\documentclass[prb,superscriptaddress,twocolumn,eqsecnum,byrevtex]{revtex4}
\usepackage{amsfonts}
\usepackage{amssymb}
\usepackage{amsmath}
\usepackage{color}
\usepackage{graphicx}
\usepackage{subcaption}
\usepackage{time}
\usepackage[colorinlistoftodos]{todonotes}
\usepackage{sidecap}
\begin{document}
%
%
\title{Energy spectrum and structure of one-dimensional few-electron Wigner crystals with and without coupling to light in cavity}

\author{Chenhang Huang}
\affiliation{Department of Physics and Astronomy, Vanderbilt University, Nashville, Tennessee, 37235, USA}
\author{Daniel Pitagora}
\affiliation{Department of Physics and Astronomy, Vanderbilt University, Nashville, Tennessee, 37235, USA}
\author{Timothy Zaklama}
\affiliation{Department of Physics and Astronomy, Vanderbilt University, Nashville, Tennessee, 37235, USA}

\author{K\'alm\'an Varga}
\email{kalman.varga@vanderbilt.edu}
\affiliation{Department of Physics and Astronomy, Vanderbilt University, Nashville, Tennessee, 37235, USA}

\begin{abstract}
Explicitly Correlated Gaussian basis is used to calculate the energies
and wave functions of one dimensional few-electron systems in
confinement potentials created by external potentials or coupling to
light in cavity. The appearance and properties of Wigner crystal-like
structures are discussed. 
It is shown that one dimensional Wigner crystals can be formed by coupling electrons to light due to the dipole self-interaction term in the light-matter Hamiltonian, provided an additional extremely weak
confining potential is present.
\end{abstract}

\maketitle

\section{Introduction}
A Wigner crystal is a solid phase of electrons, predicted by Eugene Wigner in 1934 \cite{PhysRev.46.1002}. If an electron gas has a low enough density in a uniform, neutralizing background, the system can crystallize through the formation of an electron lattice, driven by Coulomb interaction. 

Experimental study of Wigner crystals is hindered by the fact that low electron densities have to be reached in the presence of defects and impurities. Nevertheless,  Wigner crystals have been experimentally demonstrated in liquid  Helium \cite{PhysRevLett.42.795} and semiconductor hetero-structures \cite{hetero,PhysRevLett.60.2765}. These experiments have led to  intense theoretical work focusing on energetics and structures of Wigner crystals  \cite{PhysRevB.76.085341,doi:10.1063/1.4974273,Yannouleas_2007,PhysRevB.69.085116,PhysRevB.62.8108,PhysRevLett.82.5325,PhysRevB.66.165314,RevModPhys.74.1283,PhysRevB.95.115438}.

Recently, there is a renewed interest in Wigner crystals after experimentally imaging them in Moire superlattices \cite{Regan2020,PPR:PPR308353,D0CS01002B} and one-dimensional (1D) systems \cite{Deshpande2008,Shapir870,Pecker2013}. These new systems emerge as a highly conducive platform to study 
strong electronic correlations as well as topology. The most interesting experimental 
example of 1D Wigner crystals is found by real space imaging the  density
profile of electrons confined in carbon nanotubes \cite{Shapir870}. In
this experiment, the signatures of N=1 to N=6 electron crystals are clearly identified. 

In this work, we will investigate the role of the electronic correlations 
and long range Coulomb interactions in the formation of the 1D 
Wigner crystals using explicitly correlated basis functions that are especially suitable to represent the wave functions of few-electron systems. One-dimensional Wigner crystals
have been studied using the bosonisation method
\cite{PhysRevLett.71.1864}, with an effective Hamiltonian \cite{PhysRevB.95.115433} and the configuration interaction (CI) approach \cite{PhysRevB.101.125113} (see a recent review highlighting other approaches \cite{cryst11010020}). In this paper we complement these works with a more accurate approach that includes the full Coulomb Hamiltonian, using correlated basis functions to avoid the convergence issues of CI calculations, and addressing the structure of spin configurations.

The Wigner crystals are formed in external confining potentials. 
An alternative possibility  to confinement is the use of electrons interactions with cavity photons. 
The dipole self-polarization term,  ${1\over2} (\vec{\lambda}\cdot\vec{R})^2$ (where
$\vec{\lambda}$ is the interaction strength and $\vec{R}$ is the dipole moment), of the light-matter interaction Hamiltonian creates a harmonic
oscillator-like confinement. We will show that coupling a 
very weakly confined few-electron system  to light in cavity
leads to tightly localised Wigner crystal structures. Such systems have not yet been experimentally discovered, but carbon nanotubes have been studied in microwave cavities \cite{Blien2020}, and  coherent spin states in carbon nanotubes coupled to cavity photons have been investigated \cite{Cubaynes2019}. Other 1D systems confined in parabolic potentials \cite{PhysRevLett.126.253601} or 1D optical lattices in  cavity \cite{PhysRevA.83.043606} have also been  studied.

1D systems have been used as test cases mimicking more complicated dynamics because numerical solutions are easier in 1D. This interest is intensified with the investigation of light-matter coupling, where the representation of the coupled light-matter wave function requires the high dimensional product of spatial and photon bases. Restricting the nuclear or electronic motion to 1D makes model calculations 
feasible\cite{PhysRevLett.123.083201,Flick15285,Flick3026,doi:10.1021/acsphotonics.9b00648,doi:10.1063/5.0012723,Ruggenthaler2018,Schafer4883}. Our calculations
might help to improve these 1D model calculations and extend them to
more complicated cases. 

The ground state energies and wave functions will be calculated using
Explicitly Correlated 
Gaussian (ECG) basis functions \cite{RevModPhys.85.693}. The basis parameters have been optimized 
using the stochastic variational approach (SVM) \cite{suzuki1998stochastic}. The advantage of 
the approach is that the matrix elements are analytically 
available \cite{suzuki1998stochastic,doi:10.1063/1.4974273,Zaklama2019} and it produces 
very accurate energies and wave functions \cite{RevModPhys.85.693}. This method has been 
used to describe excitonic complexes 
\cite{Zhang2015,PhysRevB.93.125423,PhysRevB.61.13873,PhysRevB.69.085116,RevModPhys.85.693,PhysRevLett.83.5471} 
and two and three-dimensional quantum dots \cite{PhysRevB.59.5652,PhysRevB.63.205308}.

We will compare our results to density functional theory (DFT) 
\cite{PhysRev.136.B864,PhysRev.140.A1133} calculations. Spin-polarized
DFT calculations have often been used to analyze the structure
and energetics of two-dimensional confined electron systems and Wigner 
crystals \cite{PhysRevB.67.235307,PhysRevB.68.165337,PhysRevB.70.195310,PhysRevB.74.045313,PhysRevLett.79.1389,PhysRevB.80.165112}. 
In this work, we will investigate how well the DFT densities approximate the accurate 
few-particle results. The advantage of the DFT is that it can easily be extended for larger systems while our ECG approach scales with $N!$
due to the explicit antisymmetrization of the $N$ electron wave function, which reduces the application to small systems.

\section{Formalism}
\subsection{Few-electron system in an external confining potential in 1D} 
The Hamiltonian of an $N$ electron system interacting with a Coulomb interaction and confined in an external potential $V_c$ reads as
\begin{equation}
    H_e=-{1\over 2} \sum_{i=1}^N {\partial^2\over \partial x^2_i}+\sum_{i<j}^N V(x_i,x_j)+\sum_{i=1}^N V_c(x_i),
\end{equation}
where $x_i$ is the coordinate of the $i$th electron, and atomic units are used. Due to the singular nature of the Coulomb potential, a soft Coulomb potential will be used
\begin{equation}
    V(x_i-x_j)={1\over \sqrt{(x_i-x_j)^2+1}},
    \label{sf}
\end{equation}
and the confining potential is either quadratic $V_c(x)={1\over 2}\omega^2 x^2$, or quartic  $V_c(x)={1\over 2}\omega^2 x^4$. Similar potentials are used in Ref. \cite{PhysRevB.101.125113}.

The wave function is expanded into ECGs of the form
\begin{equation}
    \psi_k(\vec{x})={\cal A}\lbrace {\rm e}^{-{1\over 2}\sum_{i<j}^N
    \alpha_{ij}^k(x_i-x_j)^2}{\rm e}^{-\sum_{i=1}^N \beta_i^k(x_i-s_i^k)^2}\chi_S\rbrace
\label{bf}
\end{equation}
where $\vec{x}=(x_1,...,x_n)$, ${\cal A}$ is an antisymmetrizer, $\chi_S$ is the $N$ electron spin function (coupling the spin to $S$) and 
$\alpha_{ij}^k$,$\beta_i^k$ and $s_i^k$ are nonlinear parameters
($k$ stands for the $k$-th set of parameters). The 
\begin{equation}
    {\rm e}^{-\beta_i(x_i-s_i)^2}
\end{equation}
function is a Gaussian shifted into position $s_i$. By optimizing
the center $s_i$ and the width $\beta_i$, one can describe the position 
of particle $i$. The 
\begin{equation}
    {\rm e}^{-{1\over 2}\sum_{i<j}^N
    \alpha_{ij}(x_i-x_j)^2}
\end{equation}
part can be used to represent the correlation between particles $i$ and $j$. The $N$ particle wave function then can be written as
\begin{equation}
    \Psi(\vec{x})=\sum_{k=1}^K c_k\psi_k(\vec{x}),
\end{equation}
where $K$ is the dimension of the basis. The linear coefficients, $c_k$, can be determined by diagonalization, and the nonlinear ones are optimized by SVM. In the SVM, the nonlinear parameters are optimized by randomly generating  a large number of candidates and selecting the ones that 
give the lowest energy \cite{RevModPhys.85.693,suzuki1998stochastic}. The size of the basis can be increased by adding the best states one by one 
and a $K$ dimensional basis can be refined by replacing states with randomly selected better basis functions. This approach is very efficient in finding suitable parameters in high dimensional spaces.

\subsection{Few-electron system in 1D coupled to photons in cavity}
In this case the Hamiltonian is given by
\begin{equation}
    H=H_e+H_{ph}=H_e+H_p+H_{ep}+H_d.
\end{equation}
$H_{ph}$ describes the electron-photon
interaction and $H_e$ is the same electronic Hamiltonian as in the previous section. 
The electron-photon interaction can be described by using the Pauli-Fierz (PF) non-relativistic QED Hamiltonian. The PF Hamiltonian can be rigorously derived
\cite{Ruggenthaler2018,Rokaj_2018,Mandal,acs.jpcb.0c03227,PhysRevB.98.235123} by
applying the Power-Zienau-Woolley gauge transformation \cite{Zienau}, with a unitary phase transformation  on the minimal coupling ($p\cdot A$) Hamiltonian in the Coulomb gauge 
\begin{equation}
    H_{ph}={1\over 2} \sum_{\alpha=1}^M \left[
-{\partial^2\over\partial p_{\alpha}^2}+\left(\omega_{\alpha} p_{\alpha}-\lambda_{\alpha}X\right)^2
\right],
\end{equation}
where $X=\sum_{i=1}^N q_i x_i$ is the dipole operator ($q_i=-1$ is the
electron charge).
This Hamiltonian describes $M$ photon modes with elongation $p_{\alpha}$, frequency $\omega_{\alpha}$, and polarization  $\lambda_{\alpha}$. The sum can be decomposed into sum of a photonic part $H_p$, dipole self-interaction $H_d$, and
$H_{ep}$ that describes the light–matter interaction in the electric-dipole form.
The photonic part is
\begin{equation}
{H}_{p}=\sum_{\alpha=1}^{M}\left(-\frac{1}{2} \frac{\partial^{2}}{\partial p_{\alpha}^{2}}+\frac{\omega_{\alpha}^{2}}{2} p_{\alpha}^{2}\right) = 
\sum_{\alpha=1}^{M} \omega_{\alpha}\left({a}_{\alpha}^{+} \hat{a}_{\alpha}+\frac{1}{2}\right), 
\end{equation}
where  $\hat{a}_{\alpha}=\sqrt{\frac{\omega_{\alpha}}{2}}\left(p_{\alpha}-\frac{1}{\omega_{\alpha}} \frac{\partial}{\partial p_{\alpha}}\right)$ is the annihilation operator, 
and $ \hat{a}_{\alpha}^{+}=\sqrt{\frac{\omega_{\alpha}}{2}}\left(p_{\alpha}+\frac{1}{\omega_{\alpha}} \frac{\partial}{\partial p_{\alpha}}\right)$ is the creation operator. With the introduction of the creation and annihilation operators, the photon states $|n_{\alpha}\rangle$ can be generated by multiple applications of the creation operators on the vacuum state $n_{\alpha}=(\hat{a}_{\alpha}^+)^n|0\rangle$, and all other photon operations can be done by using $\hat{a}_{\alpha}$ and $\hat{a}_{\alpha}^+$. 
The interaction term is
\begin{equation}
    H_{ep}=-\sum_{\alpha=1}^M\omega_{\alpha}p_\alpha \lambda_{\alpha}X=
    -\sum_{\alpha=1}^M\sqrt{\omega_{\alpha}\over 2}(\hat{a}_{\alpha}+\hat{a}_{\alpha}^+)\lambda_{\alpha}X.
\label{epi}
\end{equation}
Note that $\hat{a}_{\alpha}$ and $\hat{a}_{\alpha}^+$ only connect photon state
$|n_{\alpha}\rangle$ to $|n_{\alpha}\pm 1\rangle$, and the matrix elements 
of the dipole operator $X$ are only nonzero between spatial basis functions of angular momentum $l$ and $l\pm 1$. The strength of the electron-photon interaction is described by the effective coupling parameter
\begin{equation}
g_{\alpha}=\left|\lambda_{\alpha}\right| \sqrt{\frac{\omega_{\alpha}}{2}}.
\label{str}
\end{equation}
The dipole self-interaction is
\begin{equation}
{H}_{d}={1\over 2} \sum_{\alpha=1}^{M} \left(\lambda_{\alpha}  X\right)^{2},
\label{dpi}
\end{equation}
which describes how the polarization of the electrons acts back on the photon field. The importance of this term for the existence of a ground state is discussed in Ref. \cite{Rokaj_2018}.

We will only consider one photon mode and the wave function in this case 
will be defined as
\begin{equation}
    \Psi(\vec{x})=\sum_n\sum_{k=1}^{K_n} c_k\psi_k^n(\vec{x})|n\rangle, 
\label{pwf}
\end{equation}
where $\psi_k^n$ is the spatial basis function belonging to an $n$ photon state
and $|n\rangle$ is the photon state. The summation over $n$ includes 
photon states that significantly lower the energy. $K_n$ is the dimension of the
basis belonging to photon state $|n\rangle$.

The necessary matrix elements can be analytically calculated for both
the spatial and the photon parts. Note that the
basis functions in Eq. \eqref{bf} do not have definite angular momentum quantum numbers. During the optimization, the symmetry of the Hamiltonian will dictate 
the selection of basis functions with appropriate symmetry. For example, if the
Hamiltonian is spherically symmetric (which is not true in the present case due
to the interaction with the photons), then the wave function converges to $L=0$
angular momentum for the lowest state. In principle, one can use Wigner rotation
matrices to project out good angular momentum functions, but in our present case
many angular momentum states will be coupled with the photons and we will let 
SVM to select the proper ground state. 

\subsection{Density functional approach}
In DFT, the Hamiltonian is defined as
\begin{equation}
    H_e=-{1\over 2}{d^2\over dx^2}+V_H[\rho(x)]+V_{ex}[\rho(x)]+V_c(x),
\end{equation}
where $V_{ex}$ is the exchange-correlation potential and $V_H$ is the Hartree potential. The local density approximation (LDA) is used for the exchange-correlation
potential\cite{PhysRevB.23.5048} and the Hartree potential is defined as
\begin{equation}
    V_H[\rho(x)]=\int \int \rho(x')V(x-x') dx',
\end{equation}
where $V$ is the soft Coulomb potential defined in Eq. \eqref{sf}. The solution of the eigenvalue problem of the DFT Hamiltonian 
\begin{equation}
    H_e\phi_i(x)=\epsilon_i\phi_i(x)
\end{equation}
gives the Kohn-Sham orbitals and the density is calculated as
\begin{equation}
    \rho(x)=\sum_{i=1}^N\phi_i(x)^2.
\end{equation}
In this case we solve the eigenvalue equation on a numerical grid with 400 grid points and 0.1 a.u. grid spacing.

\onecolumngrid

\begin{figure}
  \begin{minipage}{0.85\textwidth}
        \includegraphics[width=0.35\textwidth]{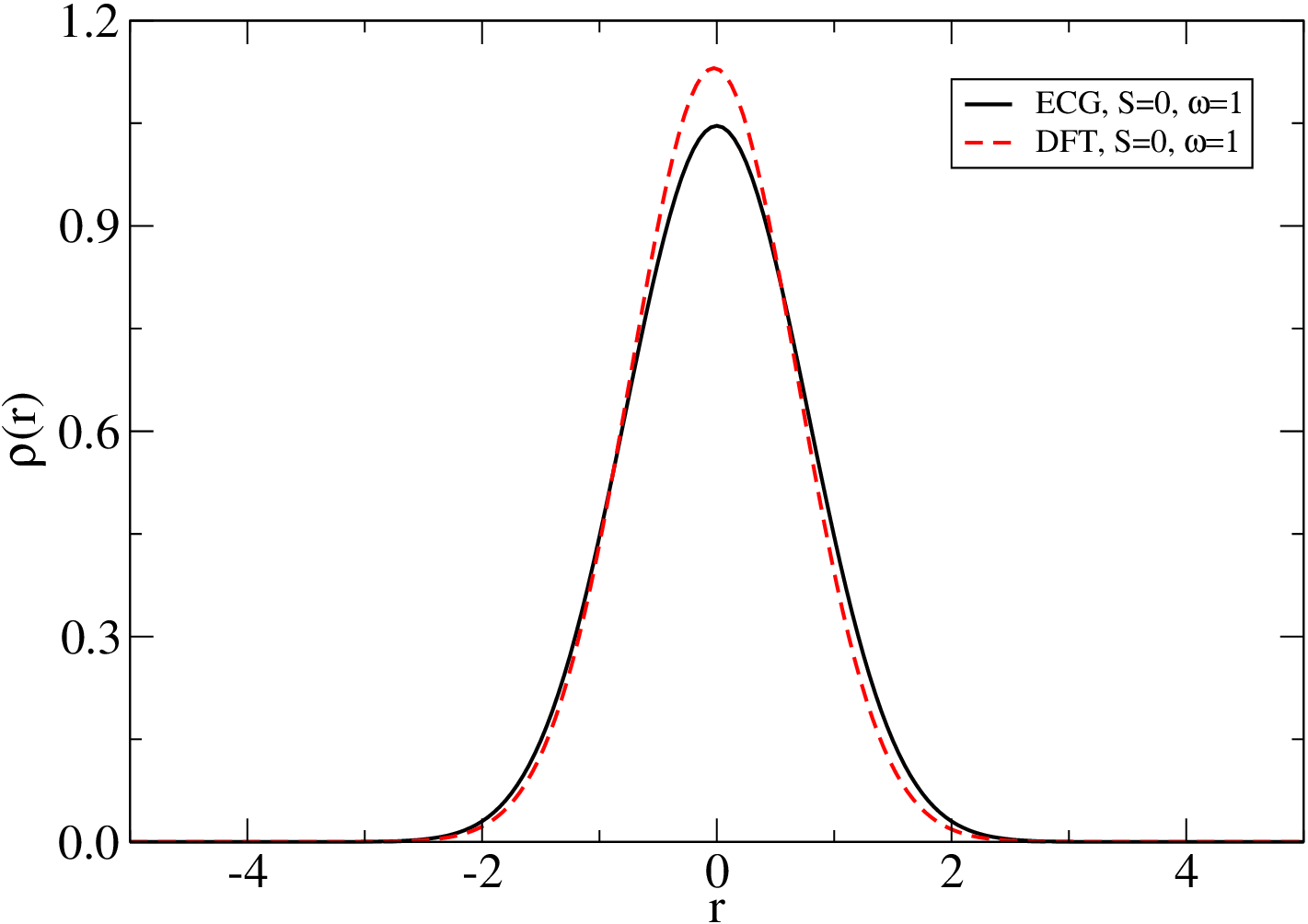}\quad
        \includegraphics[width=0.35\textwidth]{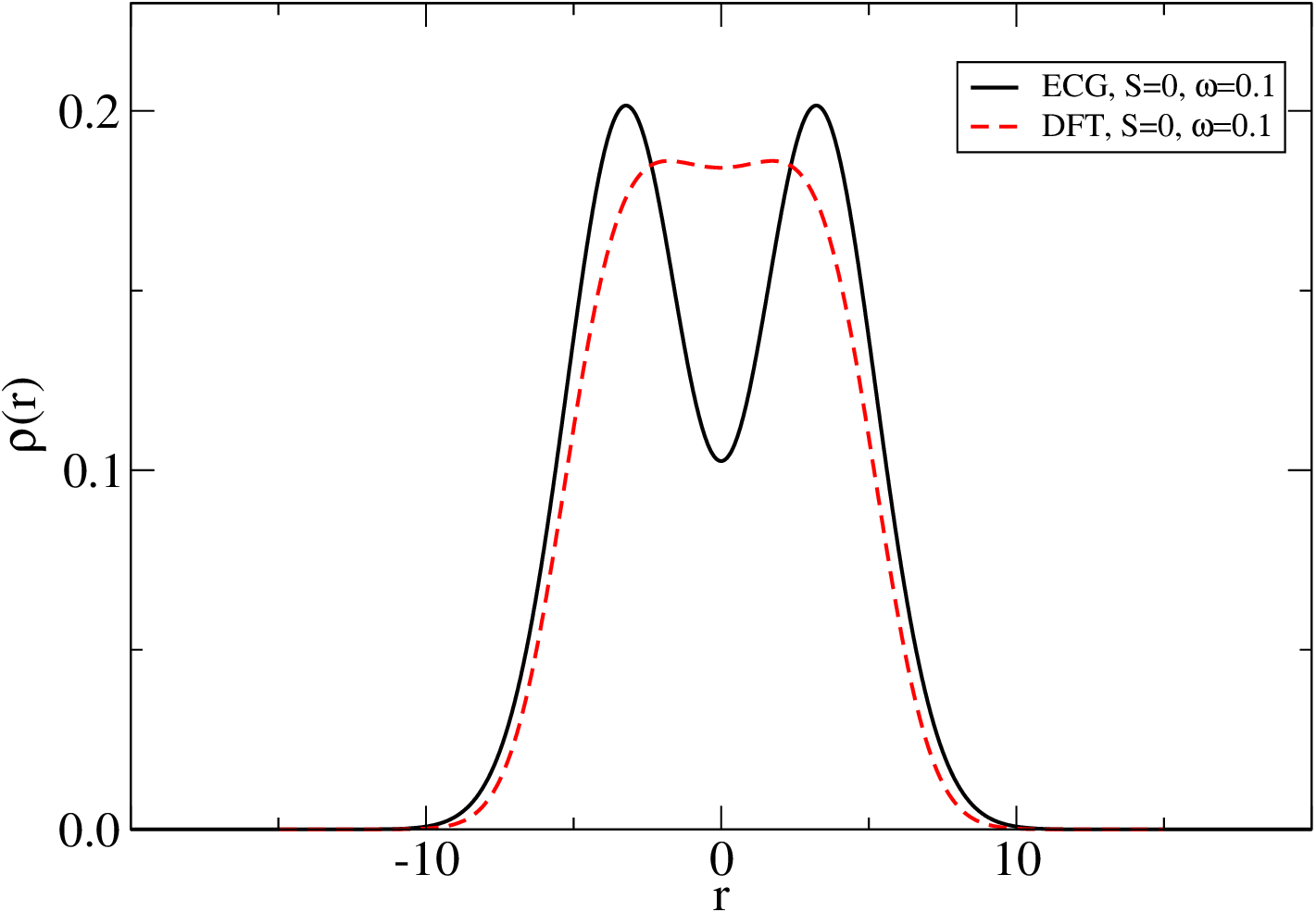}\\\vskip 0.5cm
        \includegraphics[width=0.35\textwidth]{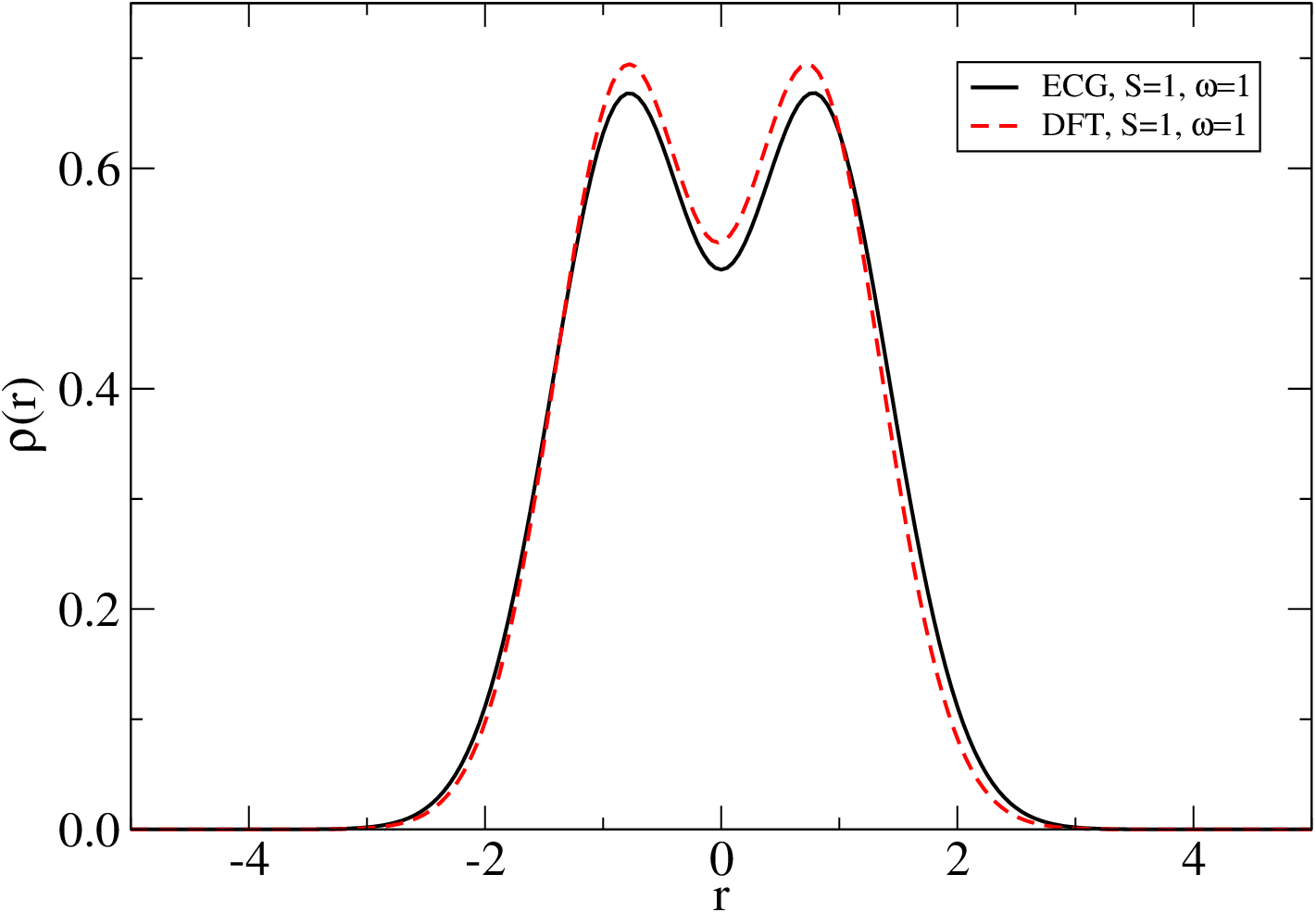}\quad
        \includegraphics[width=0.35\textwidth]{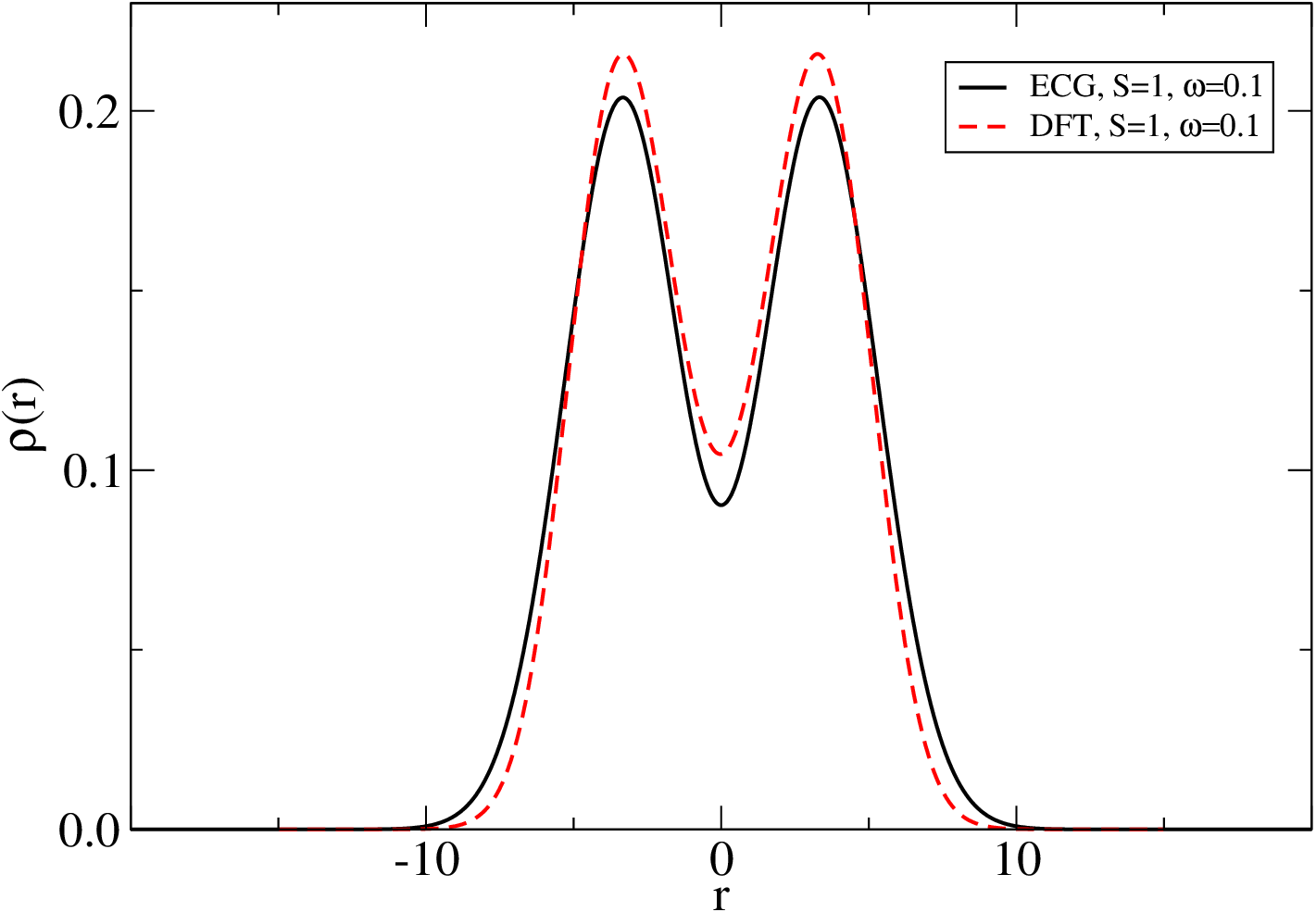}
\end{minipage}\\[1em]
  \caption{Electron density of the two electron system. Top S=0, bottom S=1, left $\omega=1$, right $\omega=0.1$ a.u. The solid curve is calculated by ECG; the dashed line is by DFT.}
\label{2ed}
\end{figure}

\vskip 1.cm

\begin{figure}
  \begin{minipage}{0.75\textwidth}
        \includegraphics[width=0.39667\textwidth]{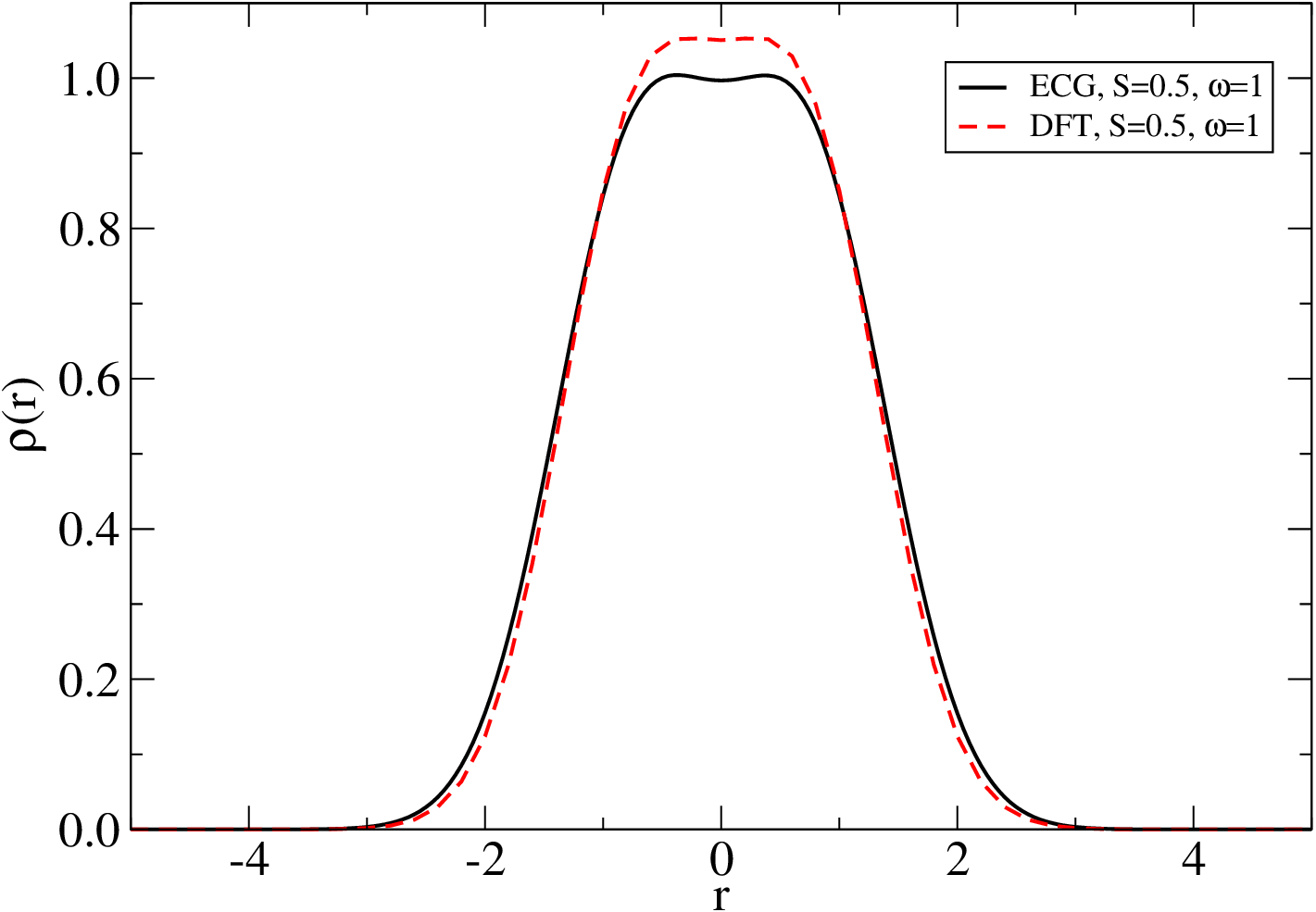}\quad
        \includegraphics[width=0.39667\textwidth]{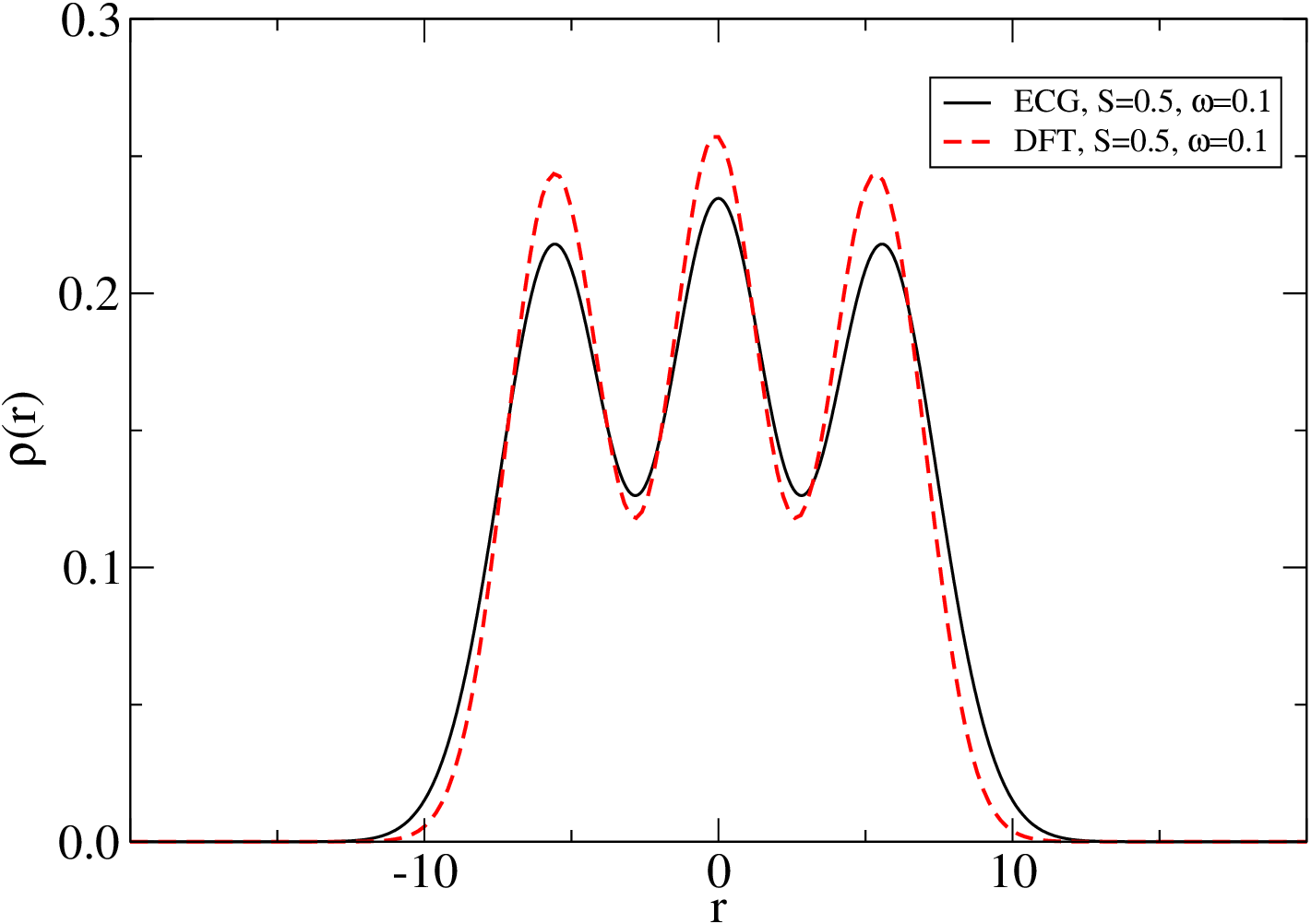}\\\vskip 0.5cm
        \includegraphics[width=0.39667\textwidth]{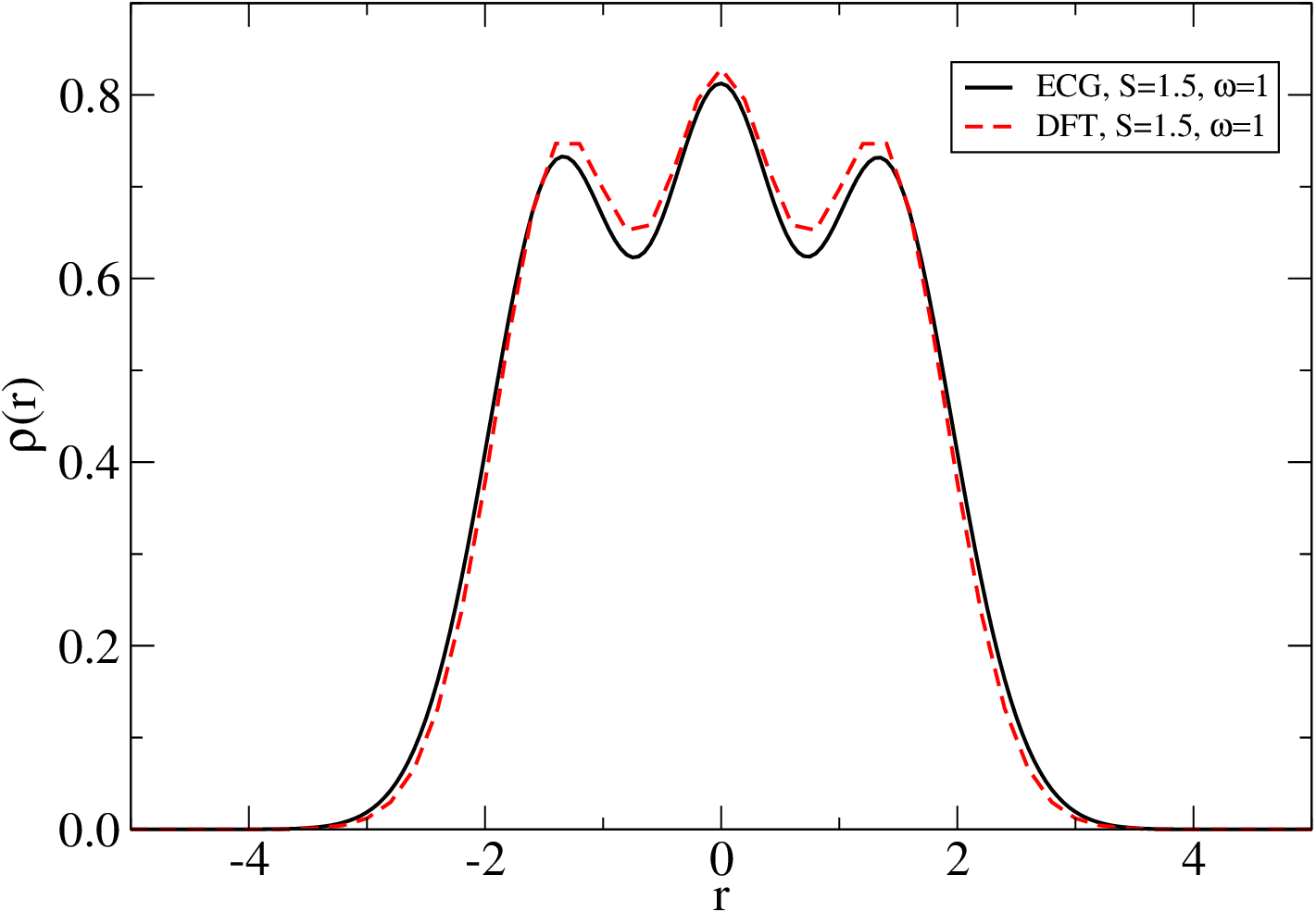}\quad
        \includegraphics[width=0.39667\textwidth]{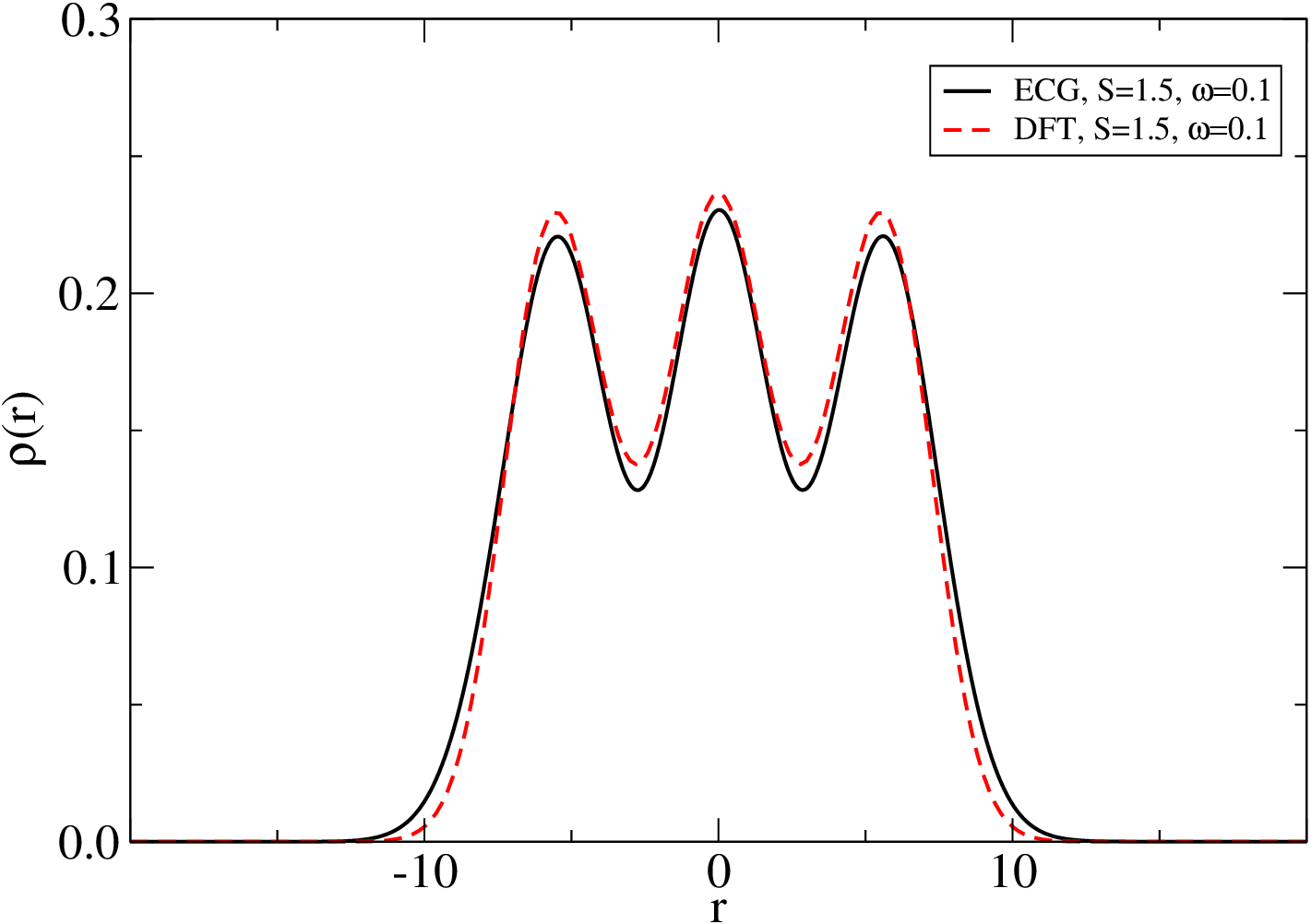}
\end{minipage}\\[0.5em]
  \caption{Electron density of the three electron system. Top S=1/2, bottom S=3/2, left $\omega=1$ a.u., right $\omega=0.1$ a.u.}
\label{3ed}The solid curve is calculated by ECG; the dashed line is by DFT.
\end{figure}
\begin{figure}
  \begin{minipage}{0.75\textwidth}
        \includegraphics[width=0.39667\textwidth]{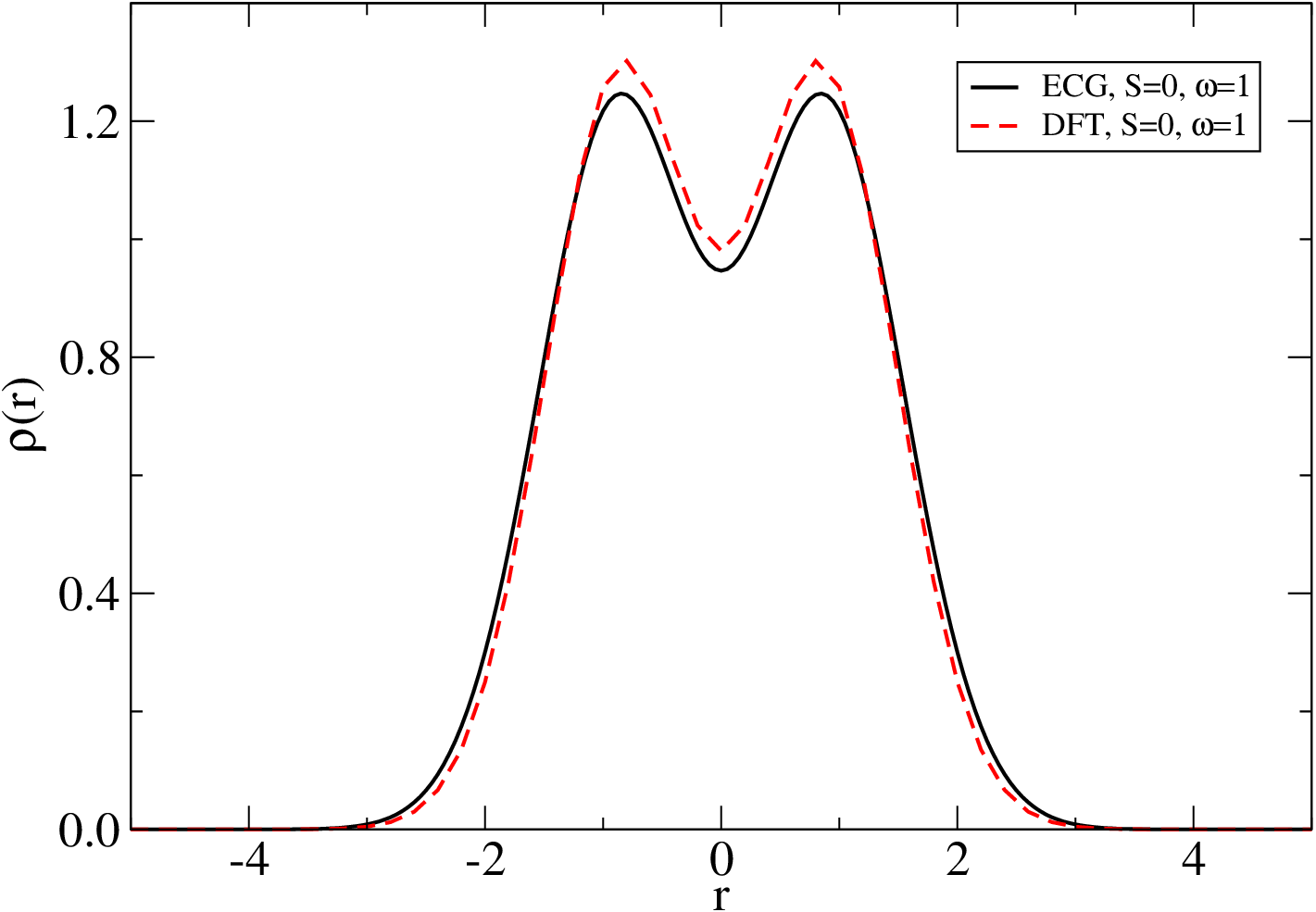}\quad
        \includegraphics[width=0.39667\textwidth]{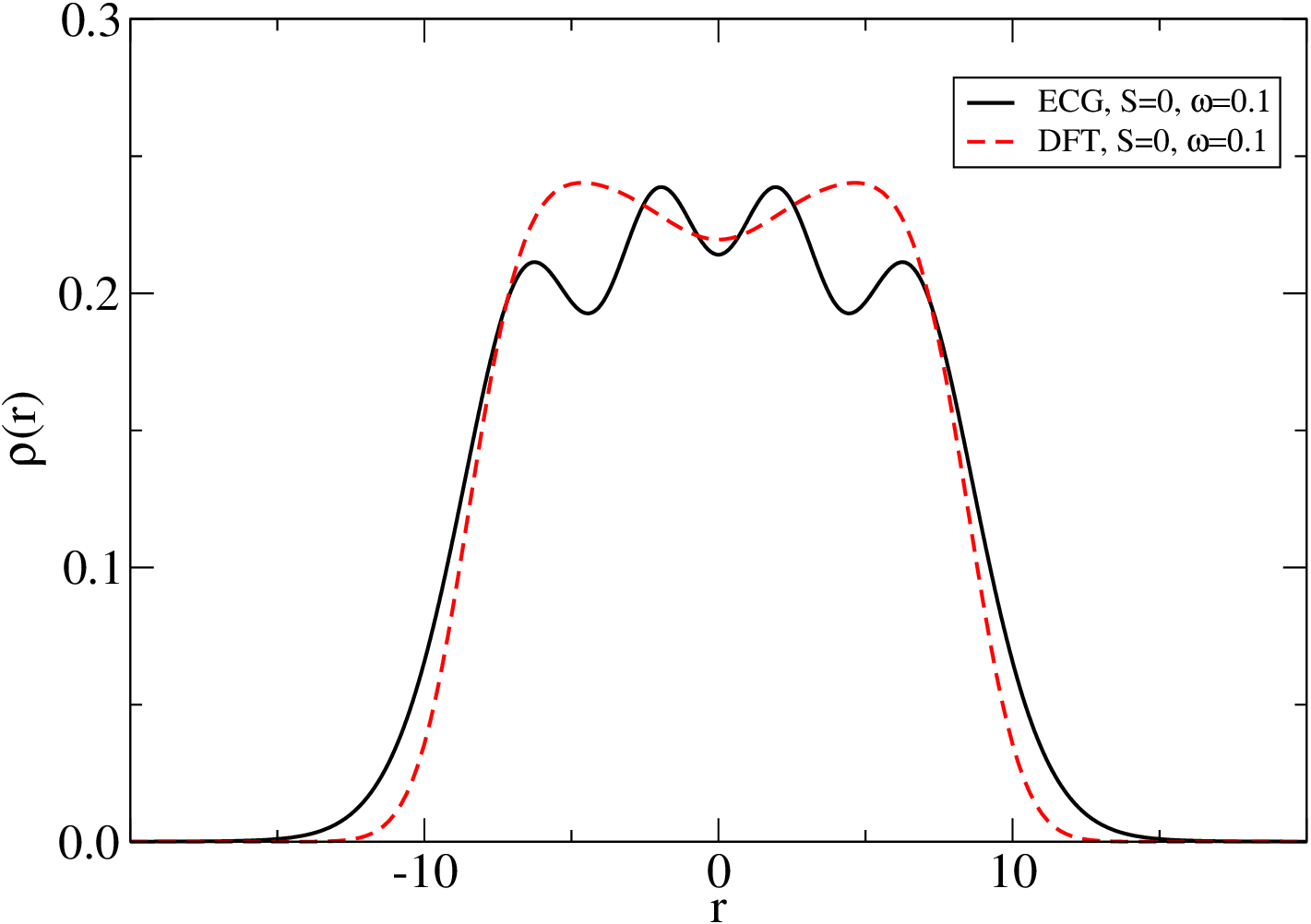}\\\vskip 0.5cm
        \includegraphics[width=0.39667\textwidth]{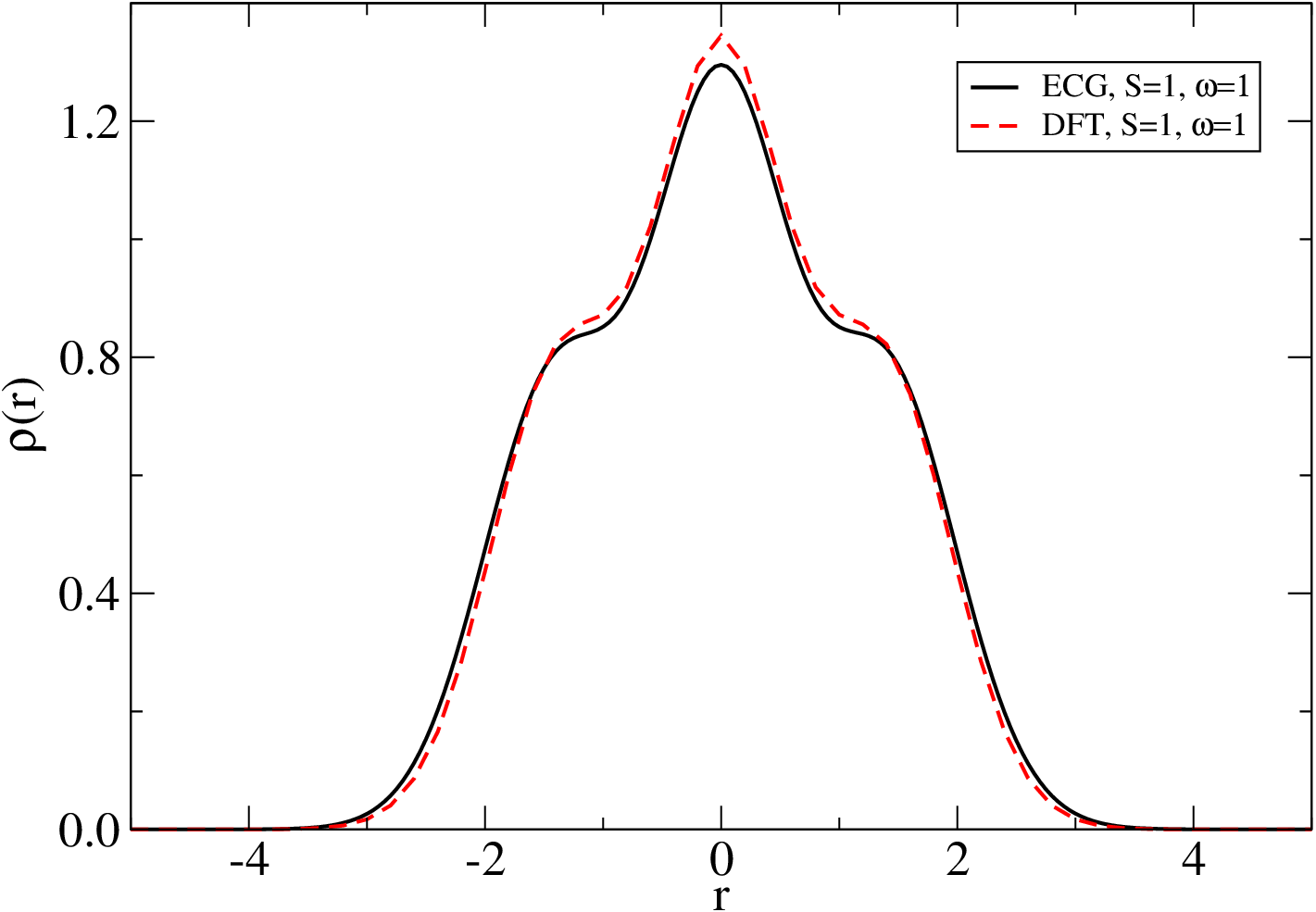}\quad
        \includegraphics[width=0.39667\textwidth]{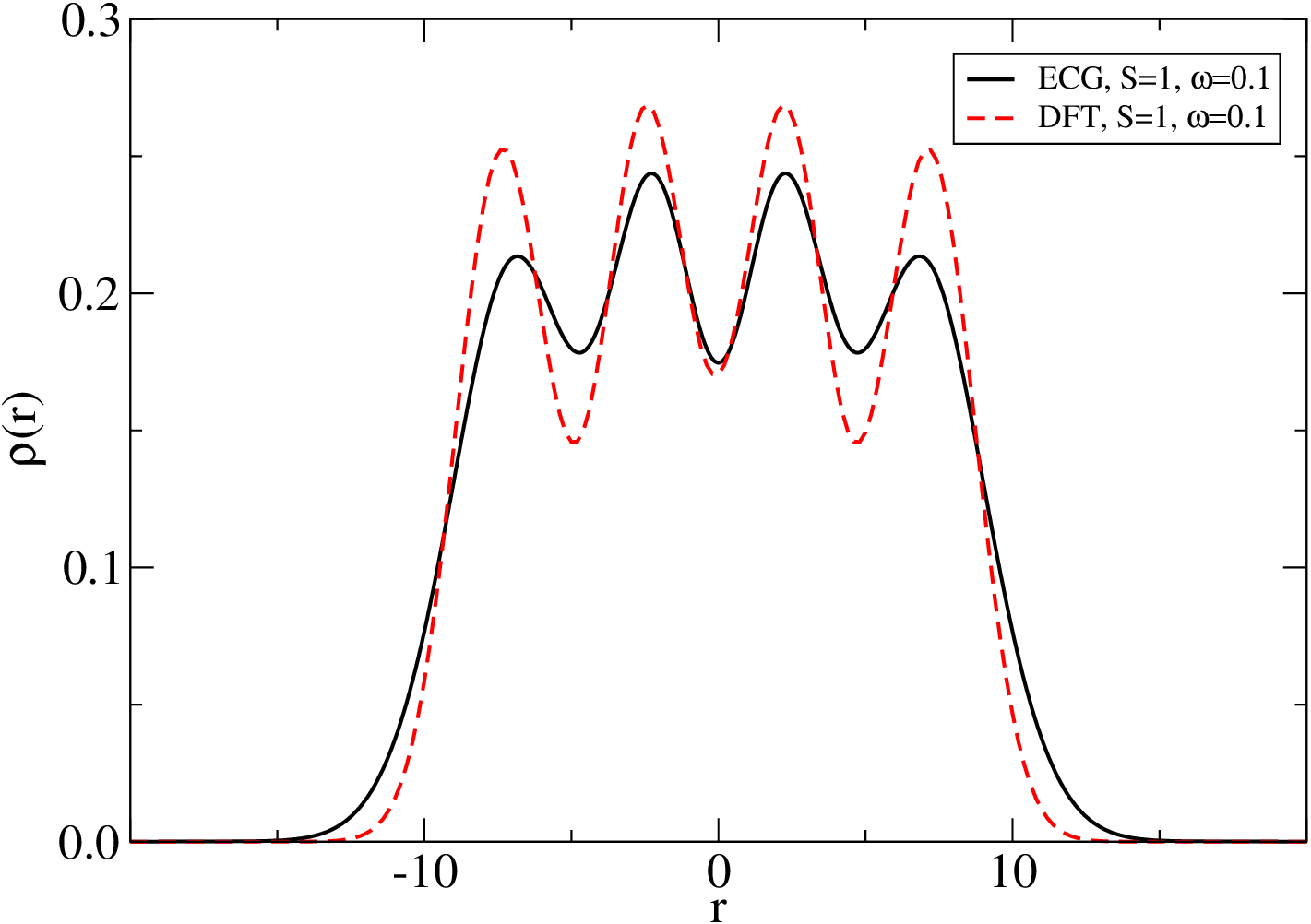}\vskip 0.5cm
        \includegraphics[width=0.39667\textwidth]{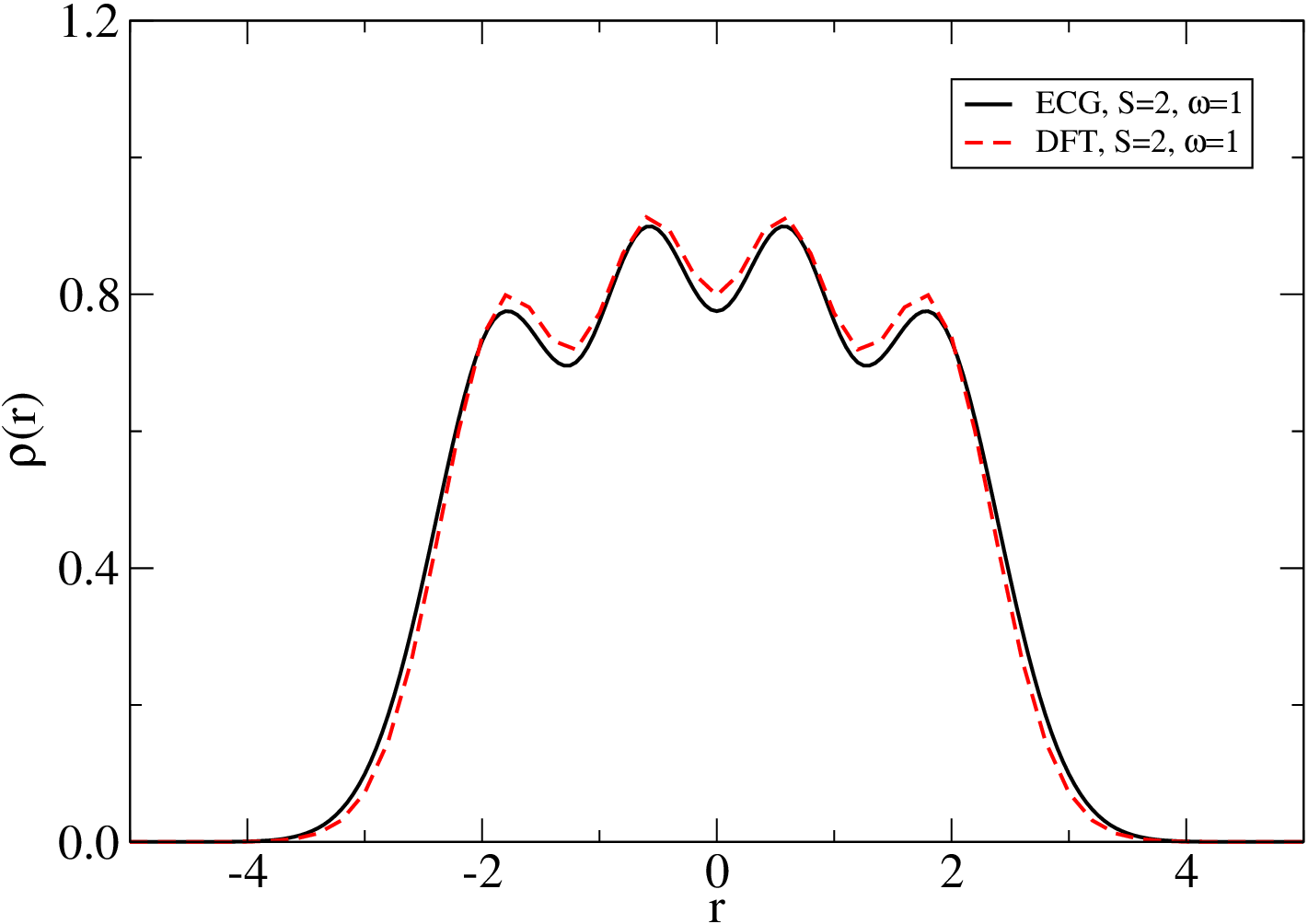}\quad
        \includegraphics[width=0.39667\textwidth]{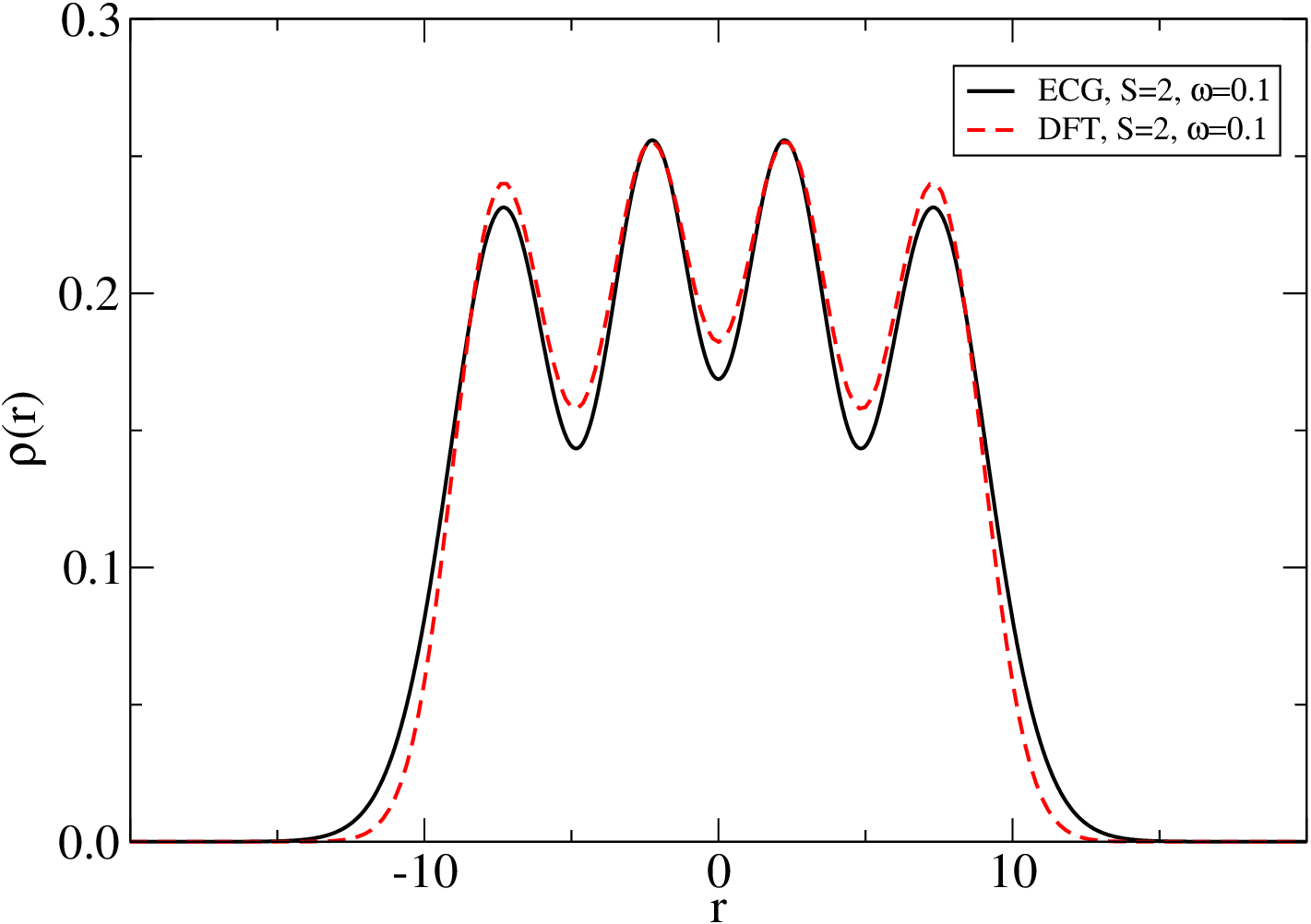}
\end{minipage}\\[1em]
  \caption{Electron density of the four electron system. Top S=0, middle S=1, bottom S=2, left $\omega=1$ a.u., right $\omega=0.1$ a.u. The solid curve is calculated by ECG; the dashed line is by DFT.}
\label{4ed}
\end{figure}

\vskip 1cm

\begin{figure}
  \begin{minipage}{0.85\textwidth}
        \includegraphics[width=0.35\textwidth]{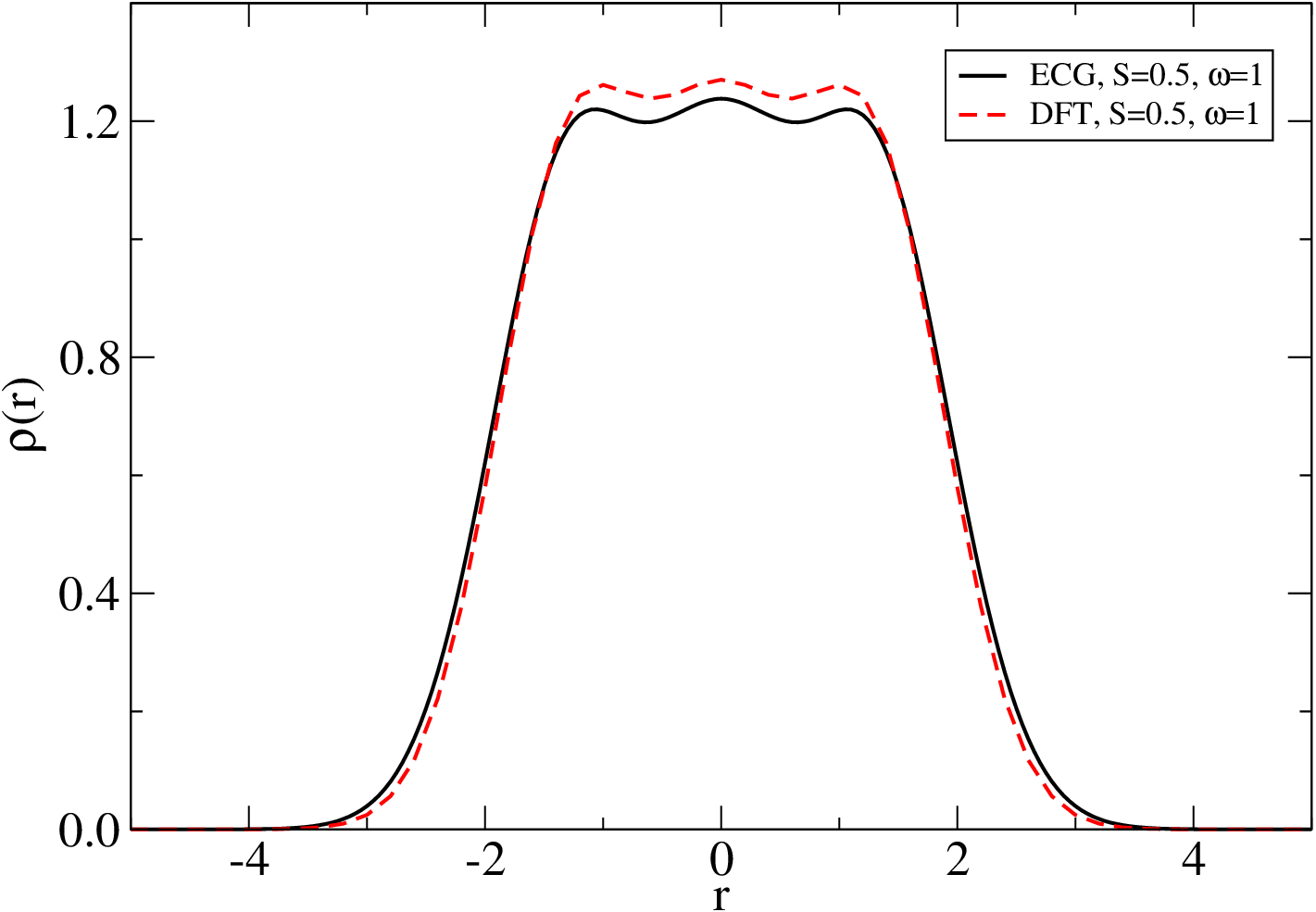}\quad
        \includegraphics[width=0.35\textwidth]{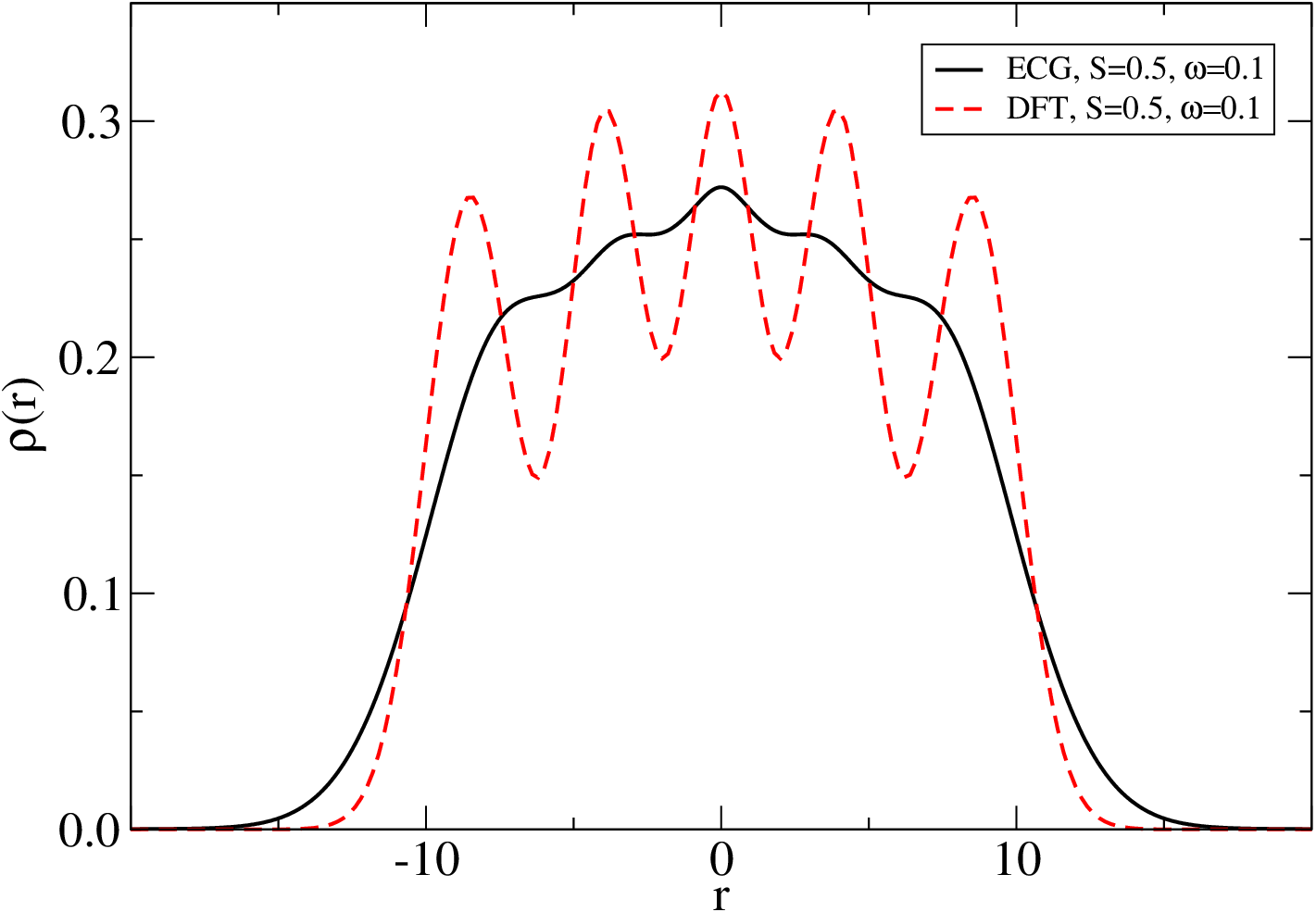}\\\vskip 1cm
        \includegraphics[width=0.35\textwidth]{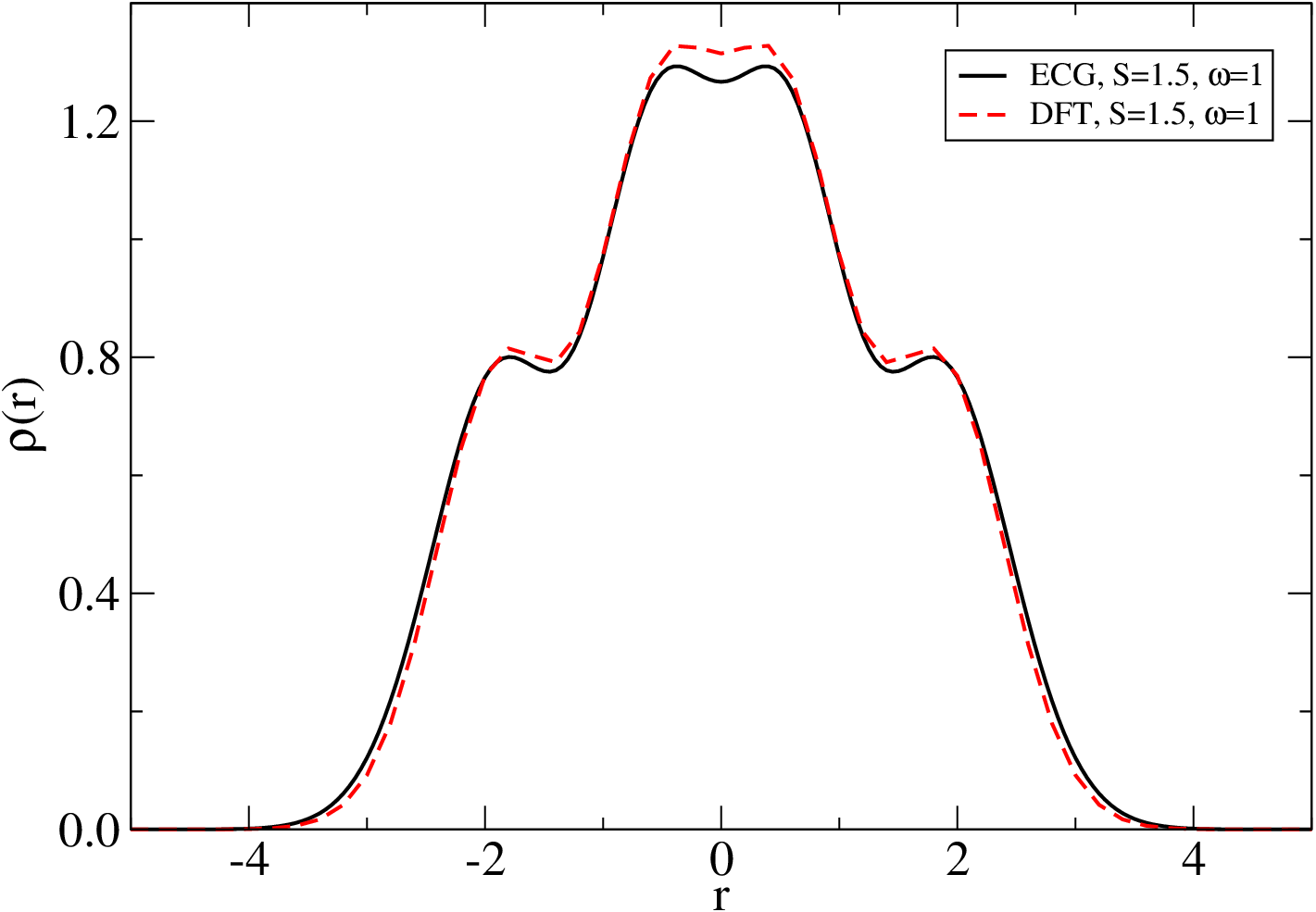}\quad
        \includegraphics[width=0.35\textwidth]{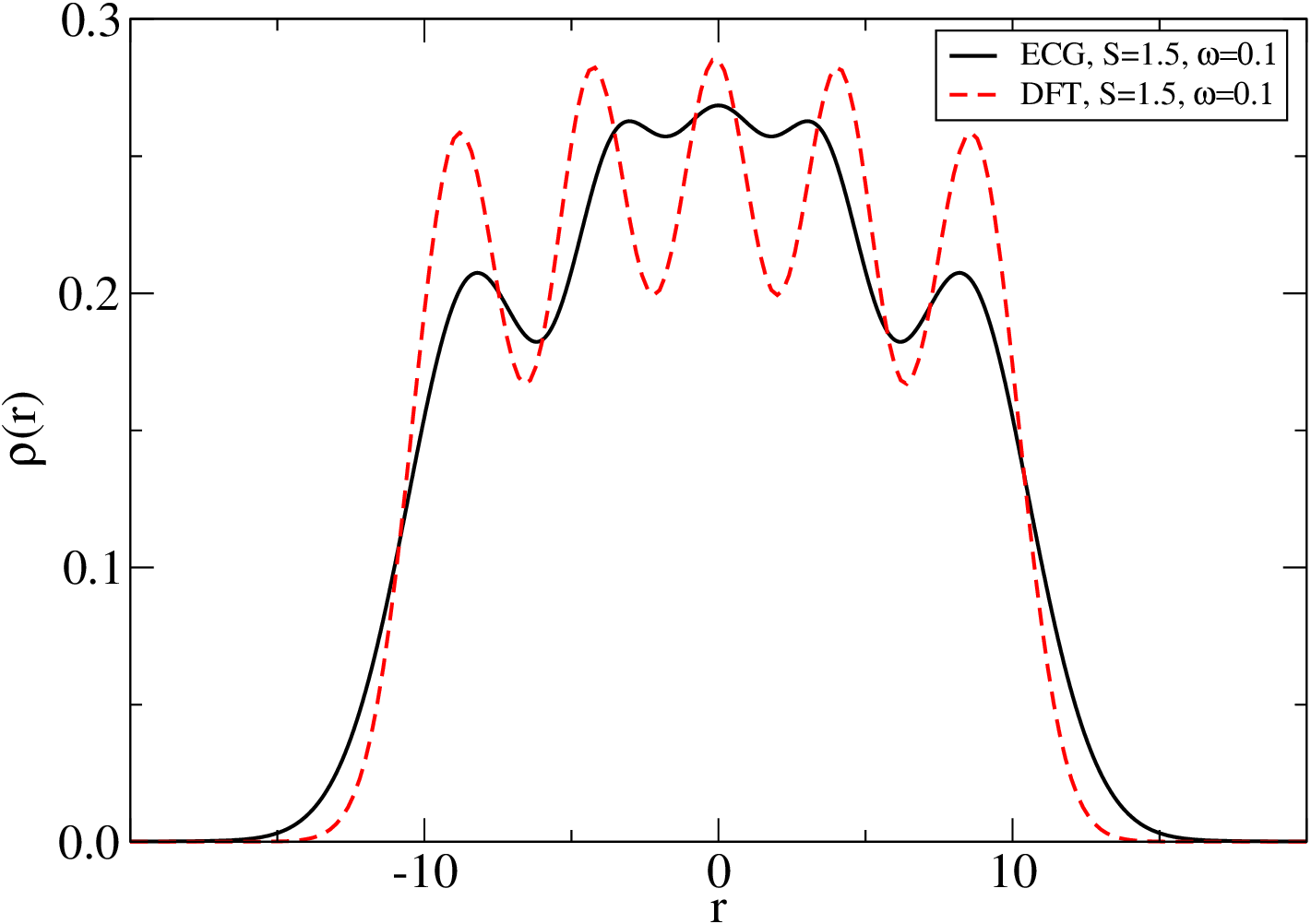}\vskip 1cm
      \includegraphics[width=0.35\textwidth]{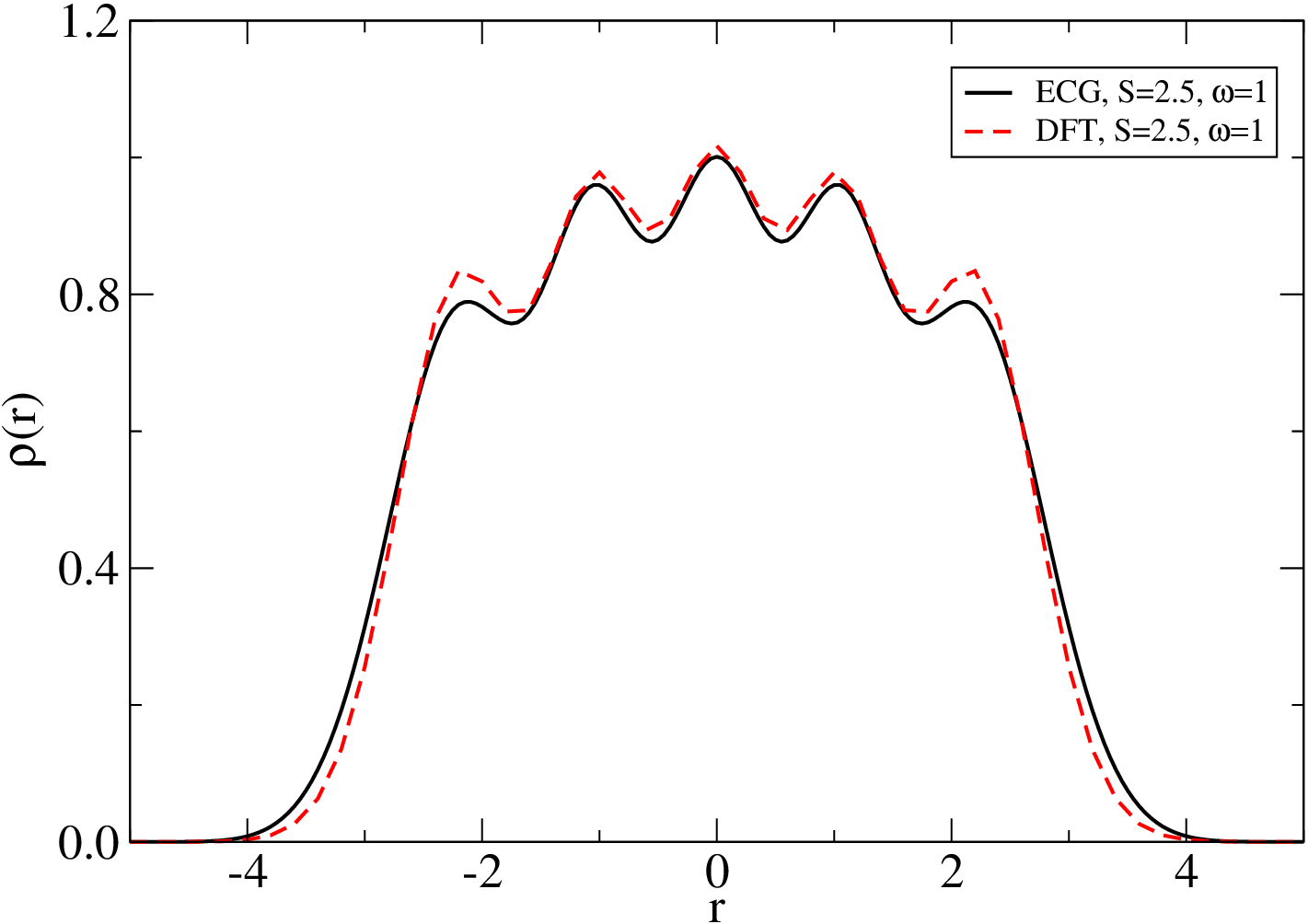}\quad
      \includegraphics[width=0.35\textwidth]{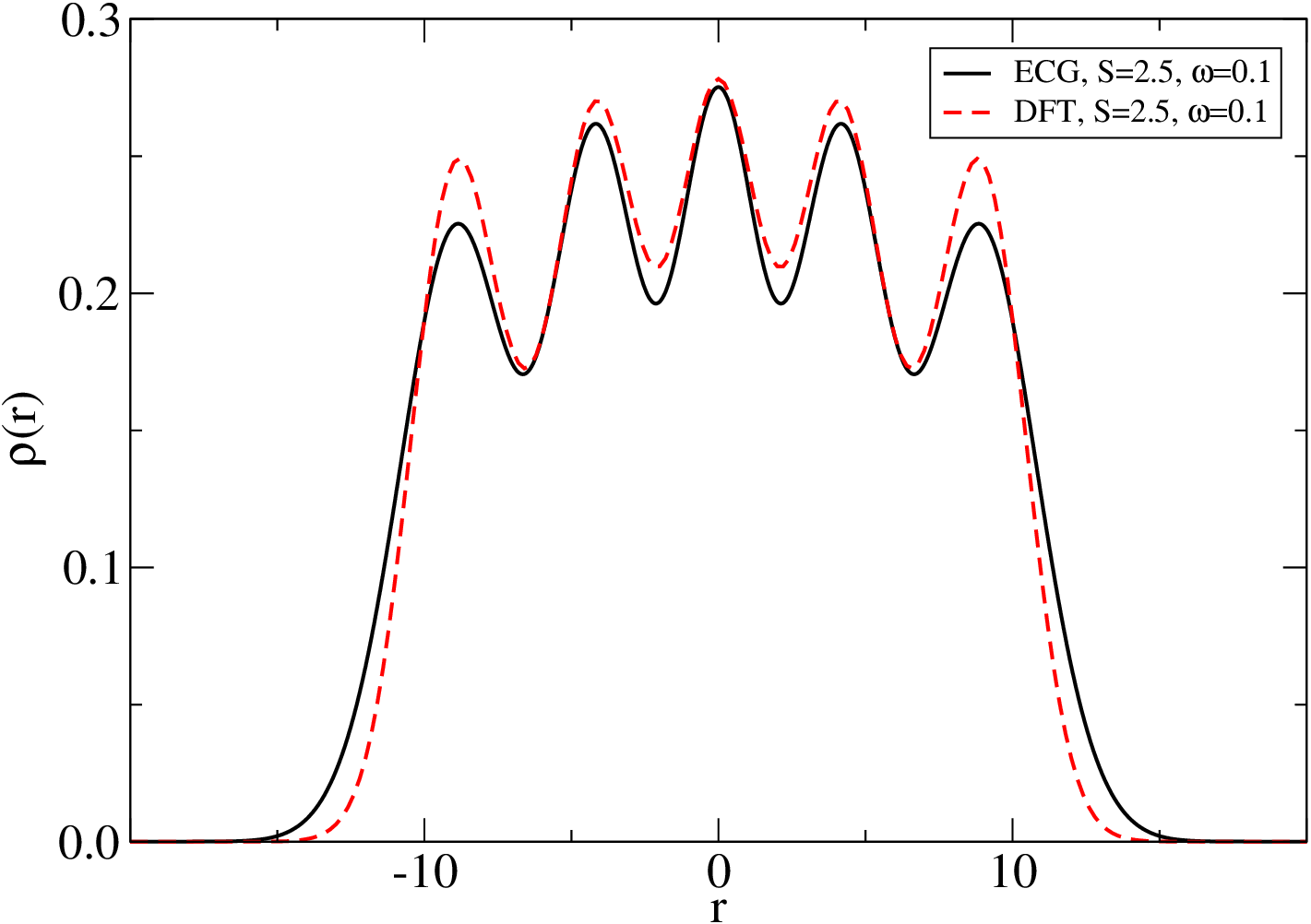}
\end{minipage}\\[1em]
  \caption{Electron density of the five electron system. Top S=1/2, middle S=3/2, bottom S=5/2, left $\omega=1$ a.u., right $\omega=0.1$ a.u. The solid curve is calculated by ECG; the dashed line is by DFT.}
\label{5ed}
\end{figure}
\vskip 1cm
\begin{figure}
  \begin{minipage}{0.85\textwidth}
        \includegraphics[width=0.35\textwidth]{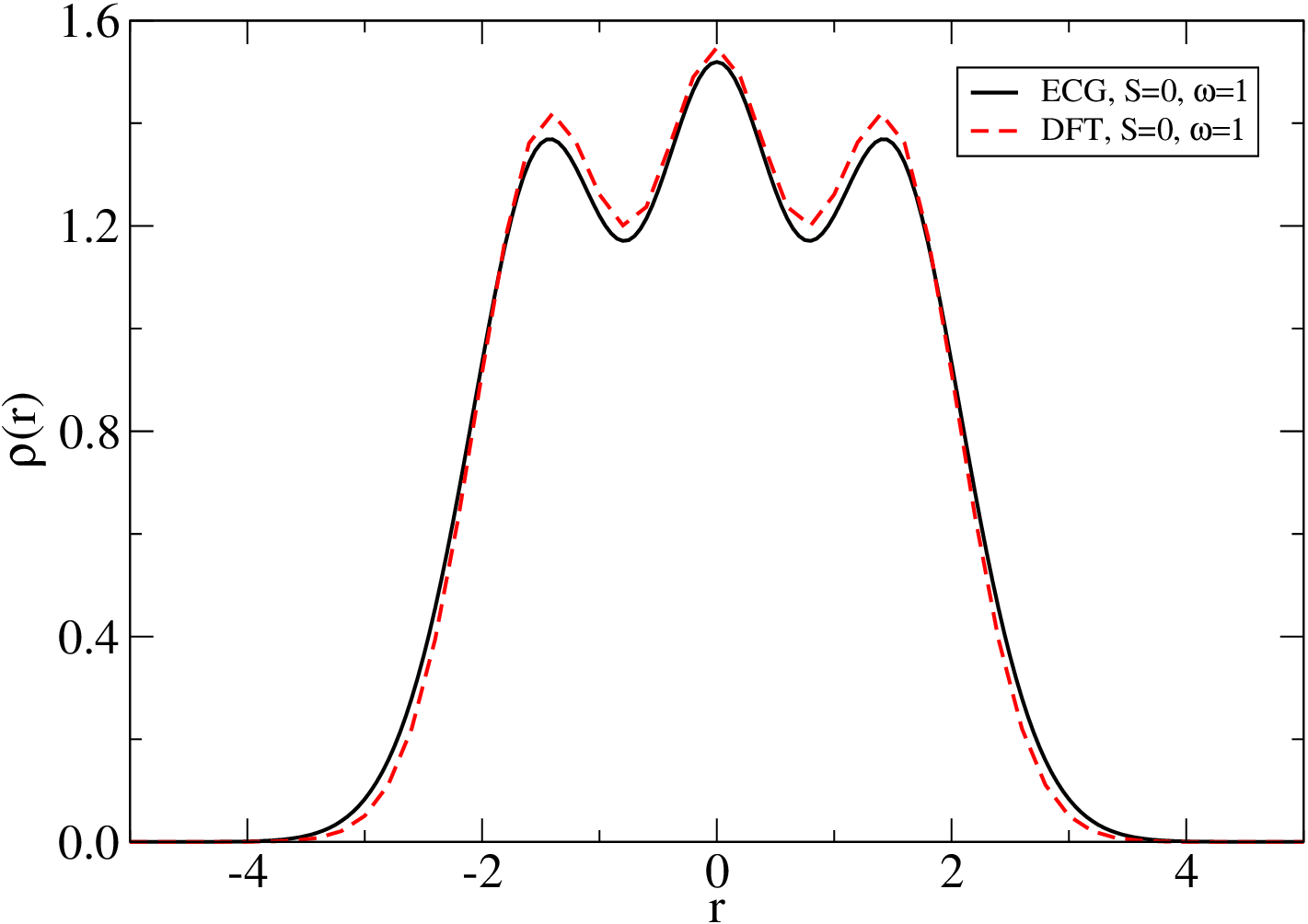}\quad
        \includegraphics[width=0.35\textwidth]{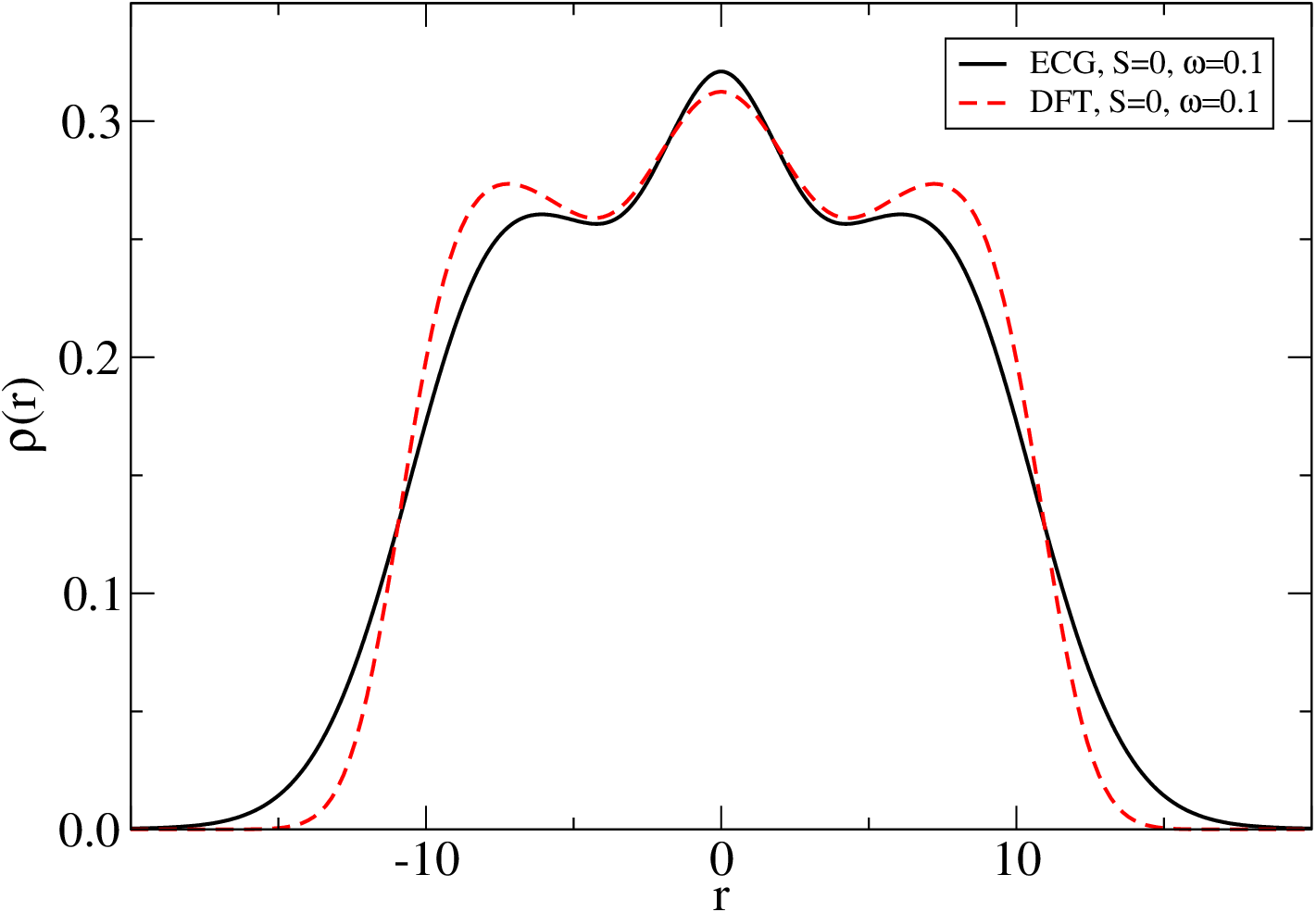}\\\vskip 1cm
        \includegraphics[width=0.35\textwidth]{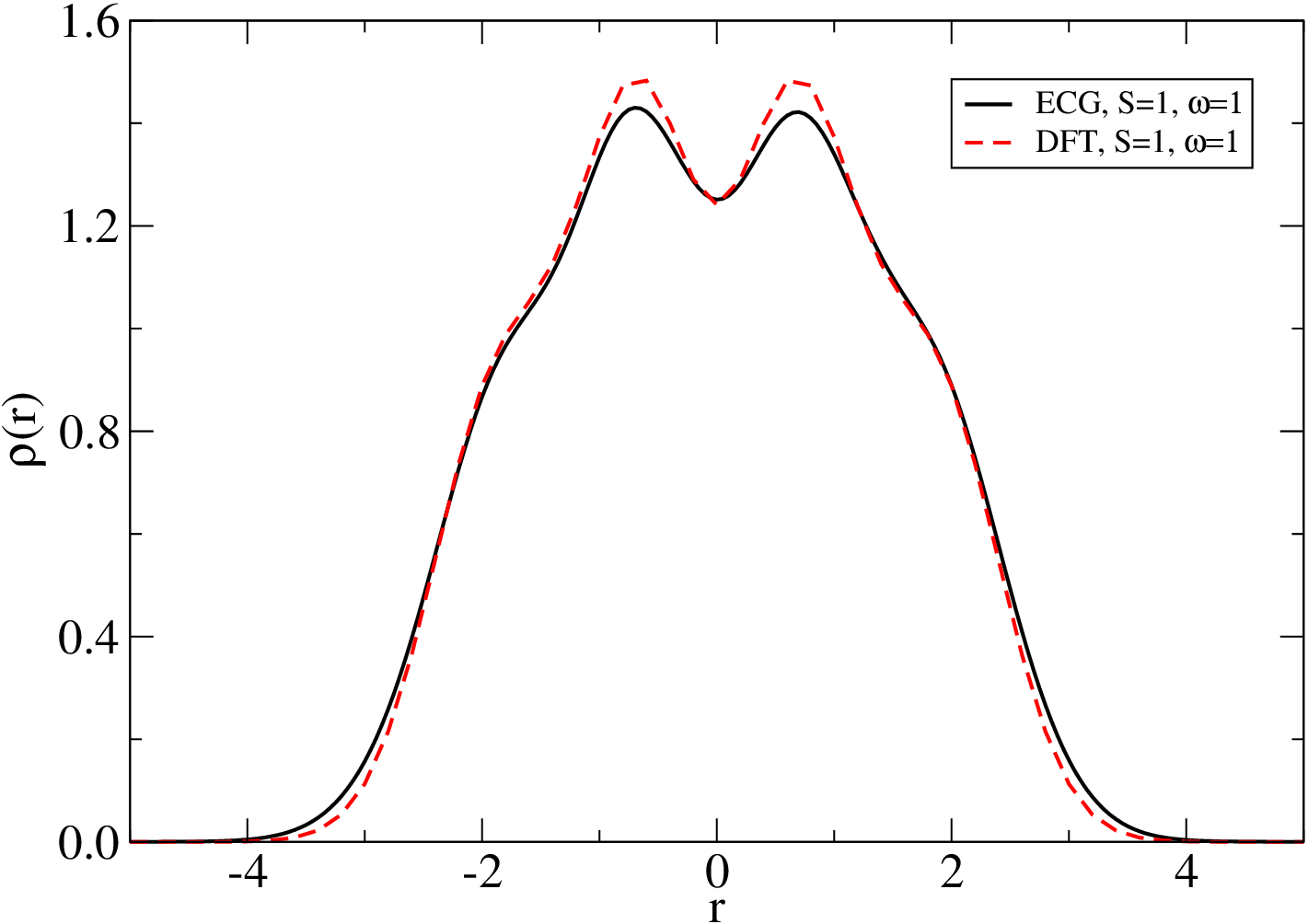}\quad
        \includegraphics[width=0.35\textwidth]{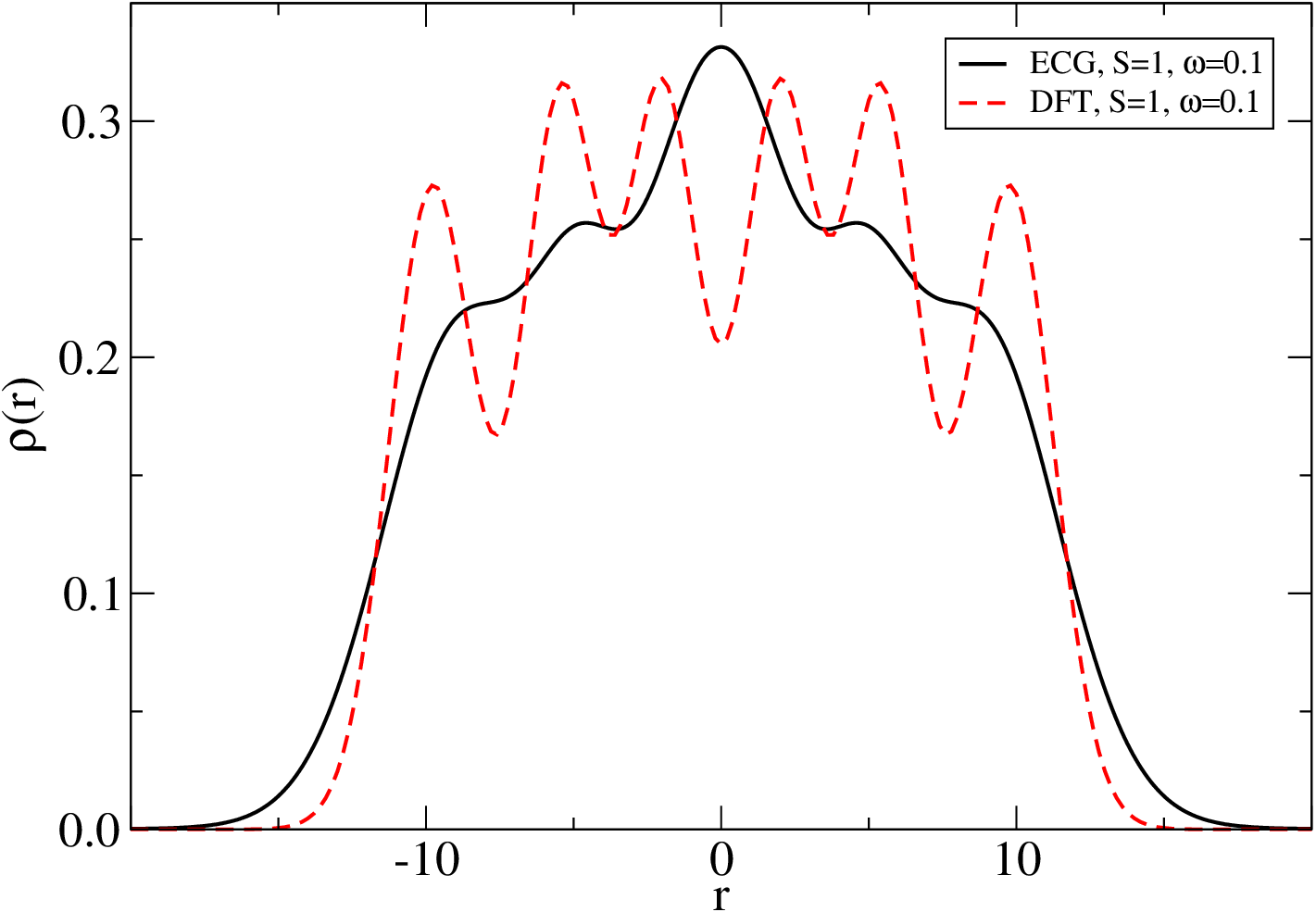}\vskip 1cm
      \includegraphics[width=0.35\textwidth]{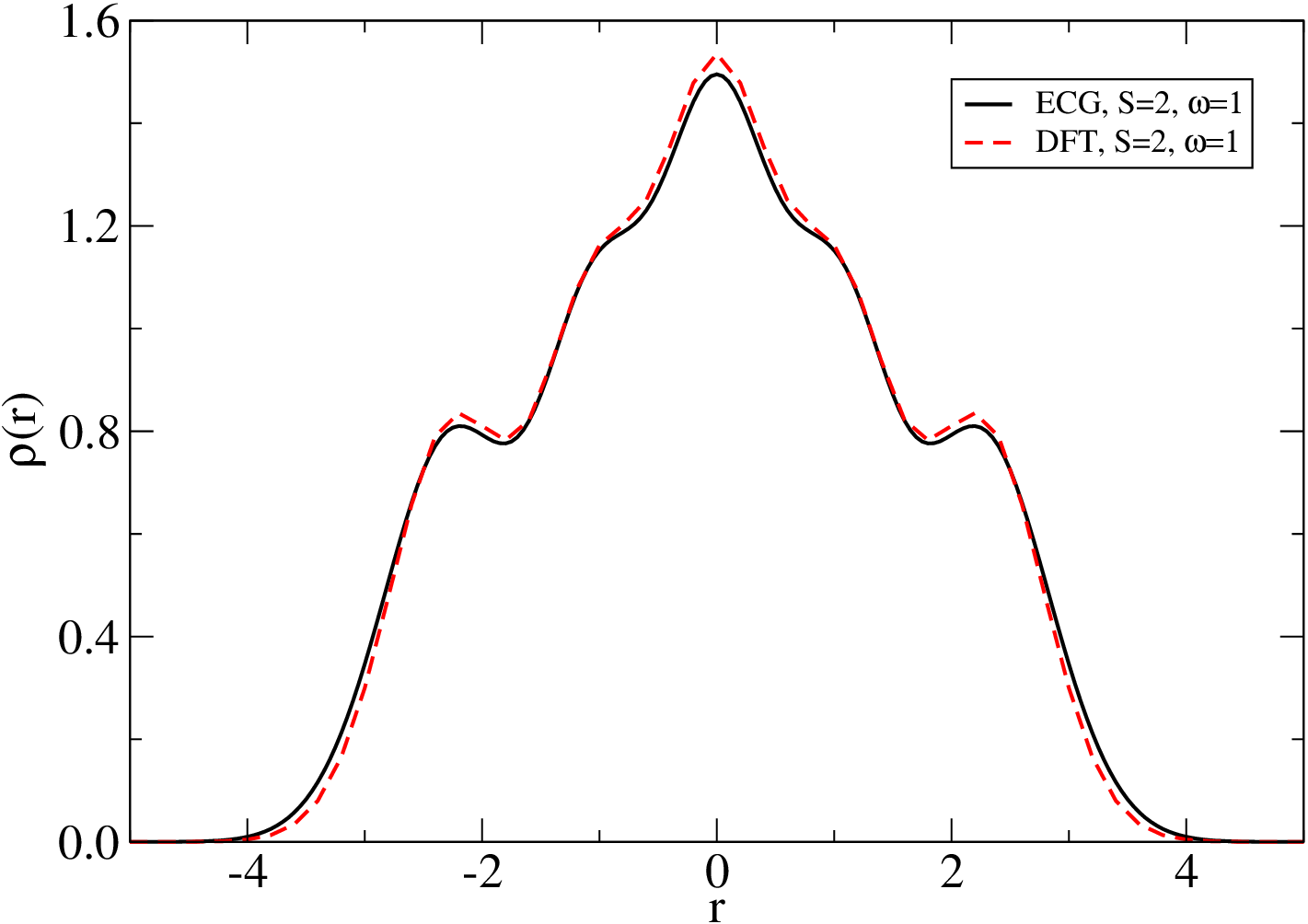}\quad
      \includegraphics[width=0.35\textwidth]{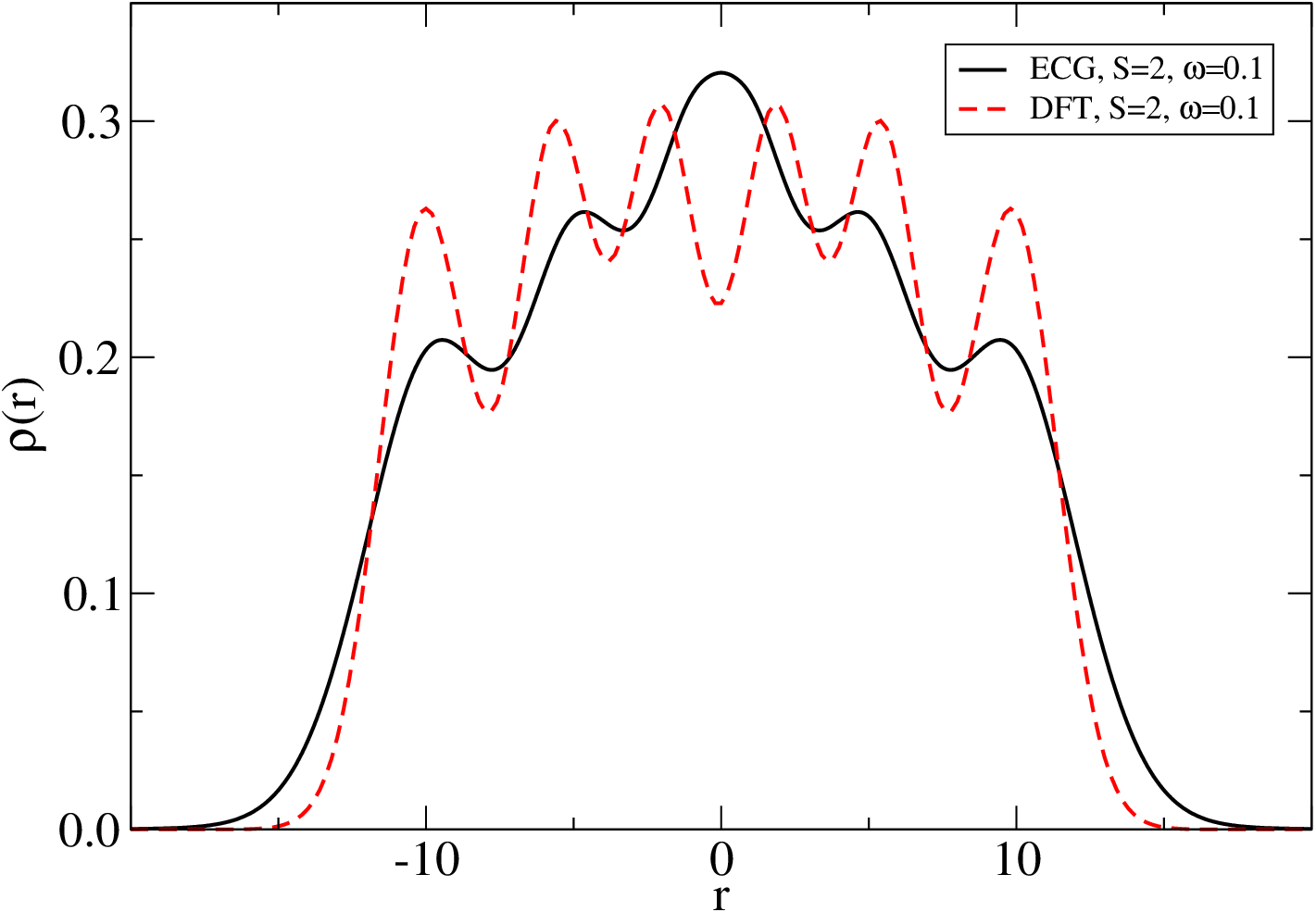}\vskip 1cm
      \includegraphics[width=0.35\textwidth]{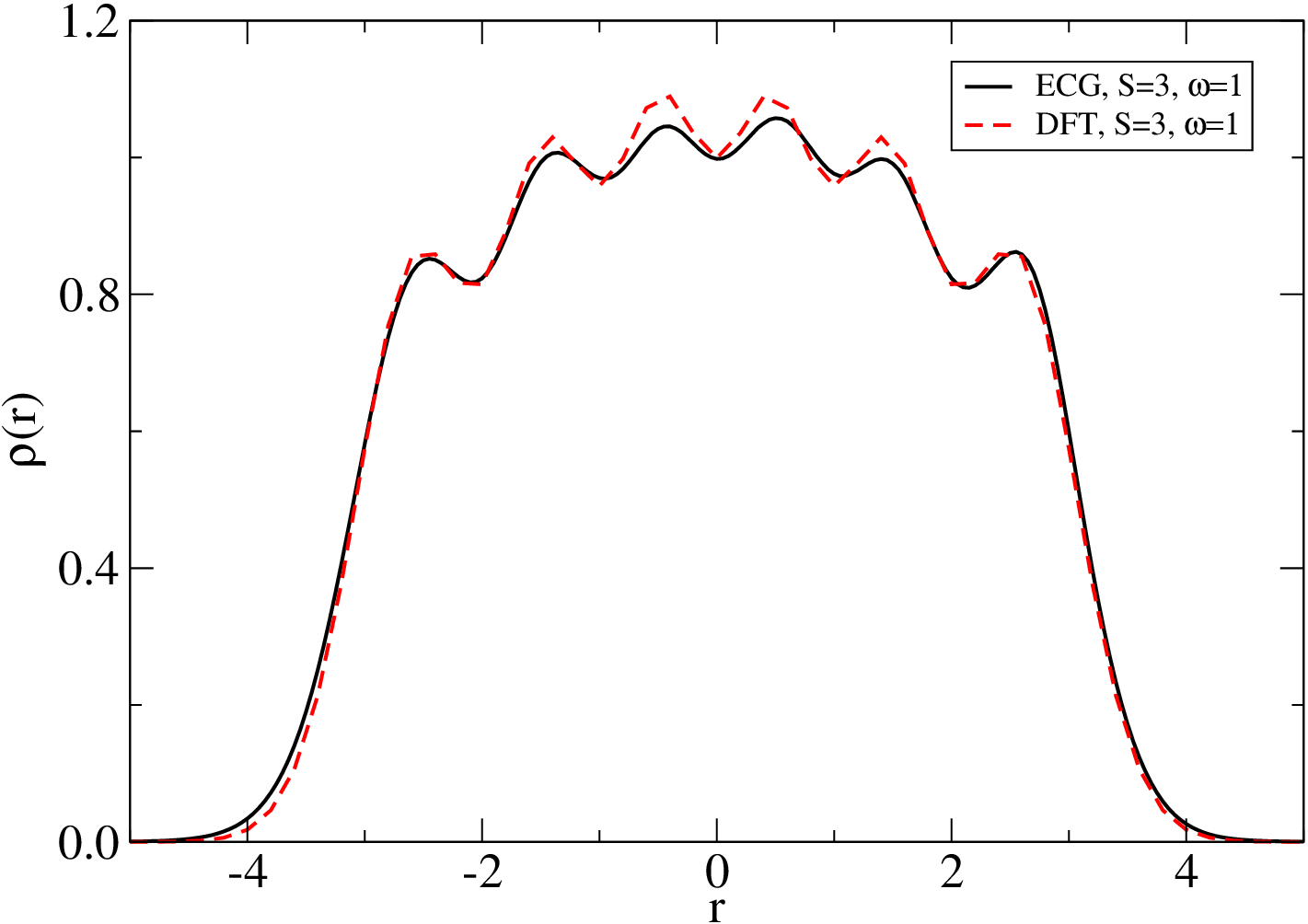}\quad
      \includegraphics[width=0.35\textwidth]{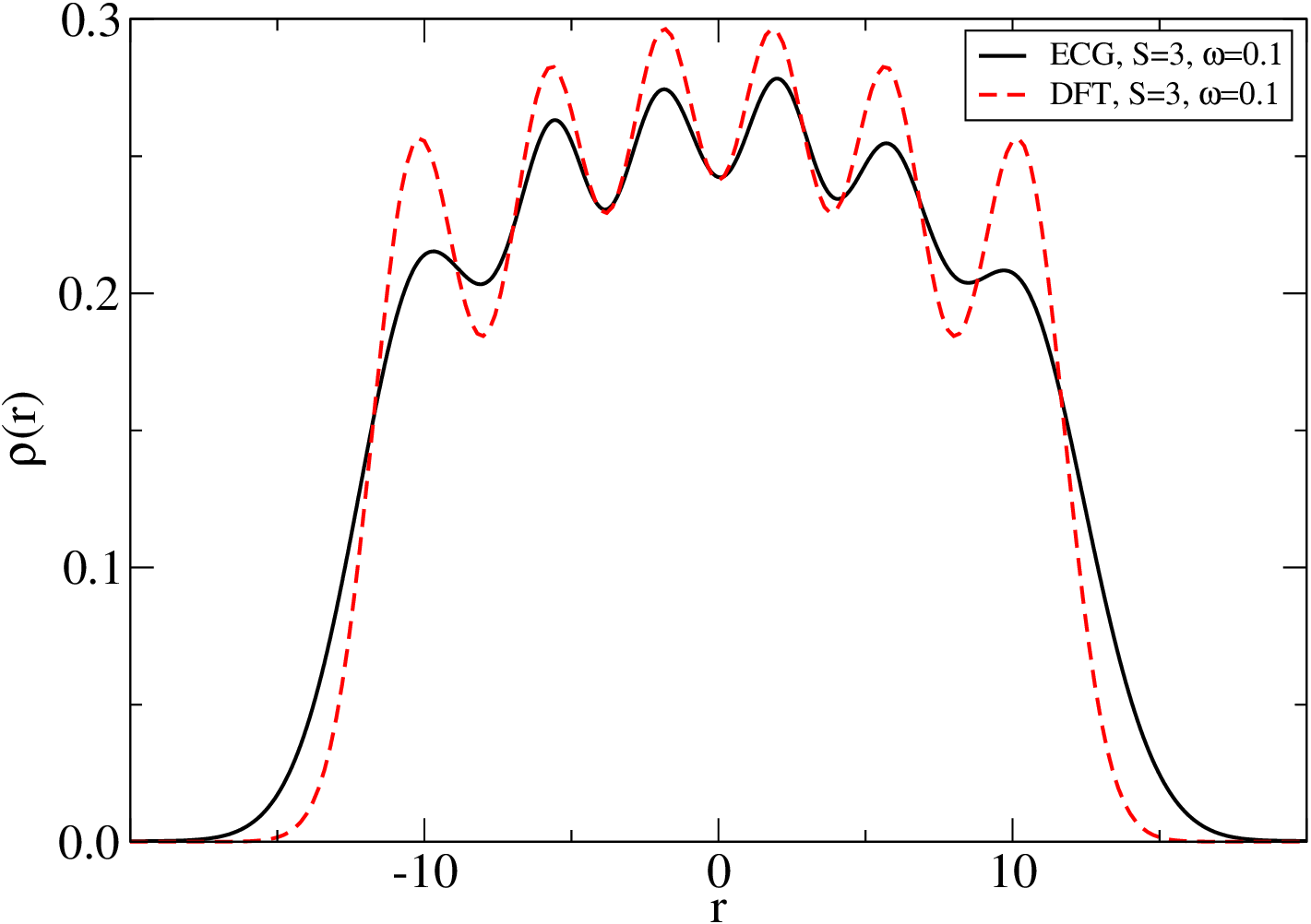}
\end{minipage}\\[1em]
\caption{Electron density of the six electron system. Top S=0, middle S=1 and S=2, bottom S=3, left $\omega=1$ a.u., right $\omega=0.1$ a.u. The solid curve is calculated by ECG; the dashed line is by DFT.}
\label{6ed}
\end{figure}
\section*{                                       }
\twocolumngrid
\section{Results and Discussion}
\subsection{Electrons in a harmonic confinement}

The DFT and ECG results are compared in Figs. \ref{2ed}-\ref{6ed} for N=2-6 particle systems
with different spin configurations. The ECG results are well converged and can be 
considered as benchmark calculations, the DFT calculations seem to provide
good approximations to the electron density in certain cases. We have tried two
different confinement potentials. The first potential, $\omega=1$ a.u., is strong 
and confines the electrons into a [-5,5] a.u. box (high electron density). The second one, $\omega=0.1$ a.u.,  confines the electrons into a [-20,20] a.u. box.
We also have calculations \cite{Supp} for $\omega=0.01$ which roughly correspond to a [-80,80] a.u. box but the results are not significantly different from the 
$\omega=0.1$ a.u. results. Besides the quadratic confinement, we have also
tested quartic confinement \cite{Supp}, but we did not observe any important 
change in the tendencies.

The two electron density (Fig. \ref{2ed}) does not show two peaks for strong 
confinement for  $S=0$, but the two peaks appear for the weaker case. In the spin-polarized $S=1$ case we have two peaks for strong and weak confinements because
the Pauli and the Coulomb repulsion together are strong enough to localize the electrons. The localization is naturally more significant in the weak confinement
case, which is shown by the increased distance and the lower density between the density peaks. The two peak structure does not disappear when the strength of the confinement increases; for $\omega=20$ a.u. the electrons are squeezed into a [-1,1] a.u. interval, but the two peaks are present in the spin-polarized case. 
The reason is simple: in the case of very strong confinement, the single particle
states of the confining potential determine the structure of the system and the 
Coulomb contribution is negligible. The two spin-polarized electrons have to occupy different orbitals, the first is the ground state and the second is the first
excited state. The ground state is node-less, the first excited state has one node and is more extended in space than the ground state. The density, the sum of the square of the two wave functions, will always have two peaks coming from the first excited state. 

The contributions of the kinetic, Coulomb, and confinement part to the total energy
are shown in Table \ref{t1}. For very strong confinement ($\omega=20$ a.u.) the
lowest single particle energy of the harmonic confinement is $E_0={1\over 2} \hbar\omega=10$ a.u., and the energy of the first excited state is $E_1={3\over 2} \hbar\omega=30$ a.u. As we have discussed above, for $S=0$ the two electrons can occupy the lowest state, and the calculated kinetic energy 9.99 a.u. and the confinement energy 10.01 a.u. (for a harmonic oscillator, the kinetic and potential energy contributions are equal according to the virial theorem)
show that this is the case. For the spin-polarized case the electrons occupy the
first two states and the energy contribution is equal to 20 a.u. for the kinetic and harmonic part. The Coulomb contribution is nearly equal in both the $S=0$ and $S=1$ cases, but it is very small compared to the kinetic and harmonic contributions. For $\omega=1$ a.u., $E_0=0.5$ a.u., and $E_1=1.5$ a.u., the single particle dominance is much less, the kinetic and confinement energy contributions are not equal to 0.5 a.u. ($S=0$) and 1 a.u. ($S=1$), and the Coulomb energy is significant compared to the other terms. For weaker confinements the Coulomb energy becomes the largest term (about half of the total energy) and the kinetic energy becomes very small.

Similar arguments are true for spin-polarized states with N=3,4,5,6 electron number cases 
shown in Figs. \ref{3ed}, \ref{4ed}, \ref{5ed} and \ref{6ed}. In particular, each spin-polarized 
case with N particles exhibits N density peaks regardless of the confinement, 
for a similar reason as in the two electron case. For example, for N=6 the first 6 states 
with increasing number of nodes are occupied, each contributing to the density. However, the 
number of density peaks does not necessarily agree with the number of particles in mixed-spin 
systems and cannot be readily determined in a simple way. As before, the electron density is 
more localized in the stronger confinements in most cases. Quartic or other forms of 
confinement do not change the shapes too much and the nodal structure is still the same 
\cite{Supp}. This 1D picture is significantly different from the 2D or 3D
cases where several single particle states are degenerate and the electrons can be placed 
in different spatial configurations. 

Both the 3 and 4 electron systems can show a single peak (see Figs. \ref{3ed}, \ref{4ed}) in mixed-spin cases 
if the confinement is strong. The two density peaks in the
$S=0$, $\omega=1$ case of $N=4$ can be easily explained. There are two electrons with up spins and two with down spins and the distinguishable particles can occupy the same spatial regions. In the $S=1$, $\omega=1$ case of $N=4$, there is one peak with two shoulders. In this case, most likely
an up-down electron pair occupies the middle region, and the two remaining electrons with spin up are on the outer region forming the shoulders.

The structure in $N=5$ and $N=6$ cases can be
understood using similar arguments to the $N=3,4$ cases. One can also think of these as a structure formed by an $N=3$ or $N=4$ system by adding two electrons. For example, the middle peak in the $S=3/2$, $\omega=1$ a.u., $N=5$ 
case is very similar to the $S=1/2$, $\omega=1$ a.u., $N=3$ density,
with two electrons added forming the outer shoulders. The same is true
for $S=2$, $\omega=1$ a.u., $N=6$ comparing with $S=1$, $\omega=1$
a.u., $N=4$.

The DFT and ECG densities are in very good agreement for $\omega=1$.
For weaker confinements the agreement is not as good, probably because
the LDA is not a good approximation for low densities where the Coulomb
interaction plays a more pronounced role. For spin-polarized systems
the DFT density remains close to the ECG even for weaker confinements.

Table \ref{t2} shows the energy of the N=2-6 systems for ECG and DFT.
Besides general trends no  agreement is expected, and the DFT with LDA is not close
to accurate ECG energies for small atoms like H, He, or Li either\cite{RevModPhys.85.693}. The general trends, however, are similar. For example, energy orders of different spin states are predicted to be the same
by ECG and DFT, especially for strong confinements. One particular
failure of DFT is the negative energy for the $N=2$, $S=1$ case and
this clearly shows that one needs to go beyond LDA. Due to the shell structure, the energies of different spin
states are very different in cases of strong confinements, but for weak confinements the energies are nearly degenerate.

\begin{table}
\begin{tabular}{|l|l|l|l|l|l|} 
 \hline
   &$\omega$ & $T$  & $V$  & $V_c$ & $E$ \\
 \hline
S=0& 0.01      & 0.007 & 0.032  & 0.025 & 0.0691 \\ 
   & 0.1       & 0.07  & 0.017  & 0.014 & 0.39   \\
   & 1.0       & 0.44  & 0.76   & 0.57  & 1.77   \\
   & 20.0      & 9.99  & 0.97   & 10.01 & 20.97  \\
 \hline
S=1& 0.01      & 0.007 & 0.032  & 0.025 & 0.0691 \\ 
   & 0.1       & 0.07  & 0.017  & 0.014 & 0.39   \\
   & 1.0       & 0.92  & 0.54   & 1.09  & 2.55   \\
   & 20.0      & 20.0  & 0.94   & 20.00 & 40.94  \\
 \hline
\end{tabular}
\caption{Energy contributions (in atomic units) for a two electron system as a function of the confinement strength. $T$ is the kinetic energy, $V$ is the Coulomb energy, $V_c$ is the confinement contribution, and $E$ is the total energy.}
\label{t1}
\end{table}

\subsection{Electrons in a cavity}
\begin{figure}
  \begin{minipage}{0.85\textwidth}
        \includegraphics[width=0.35\textwidth]{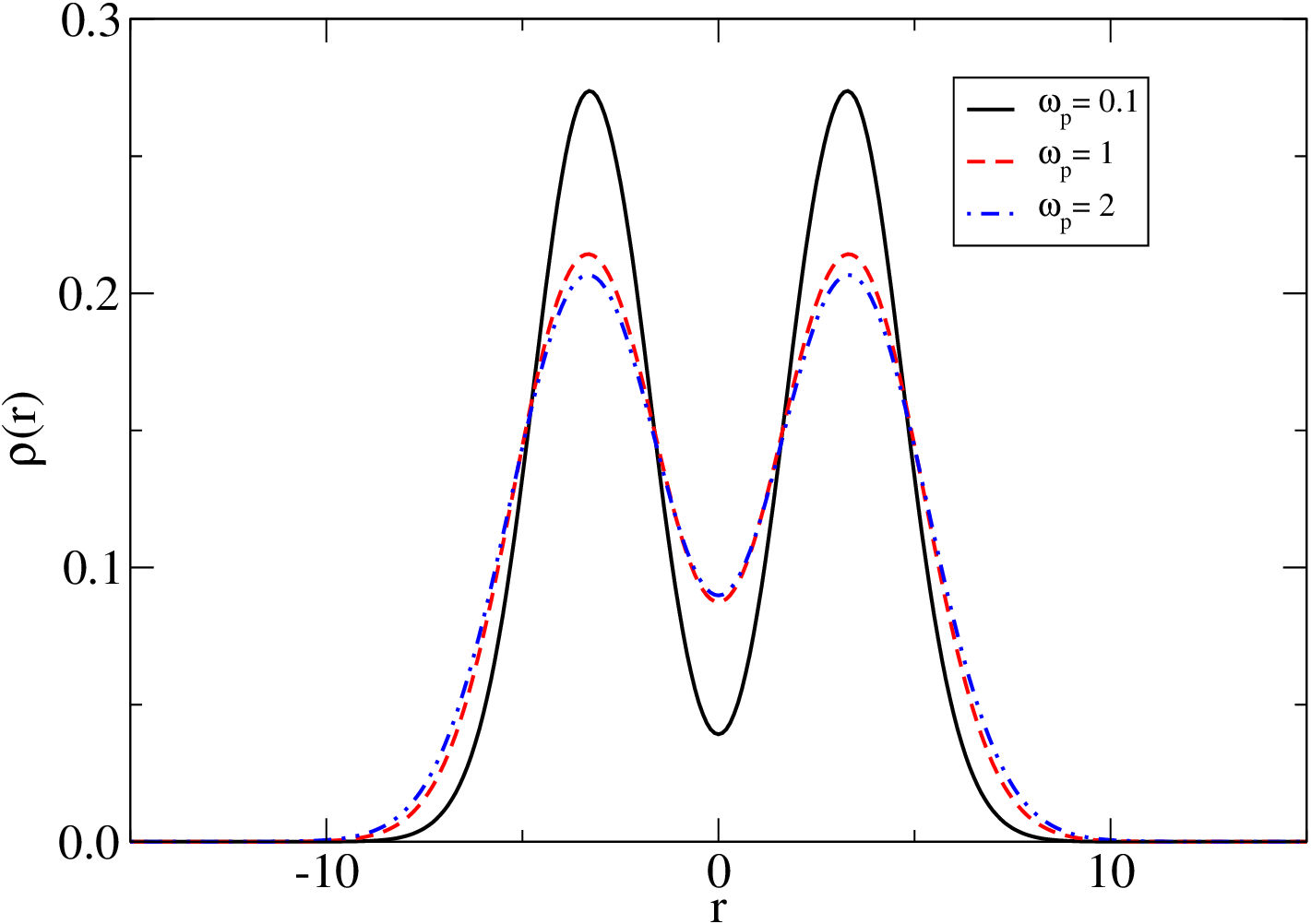}\quad
        \includegraphics[width=0.35\textwidth]{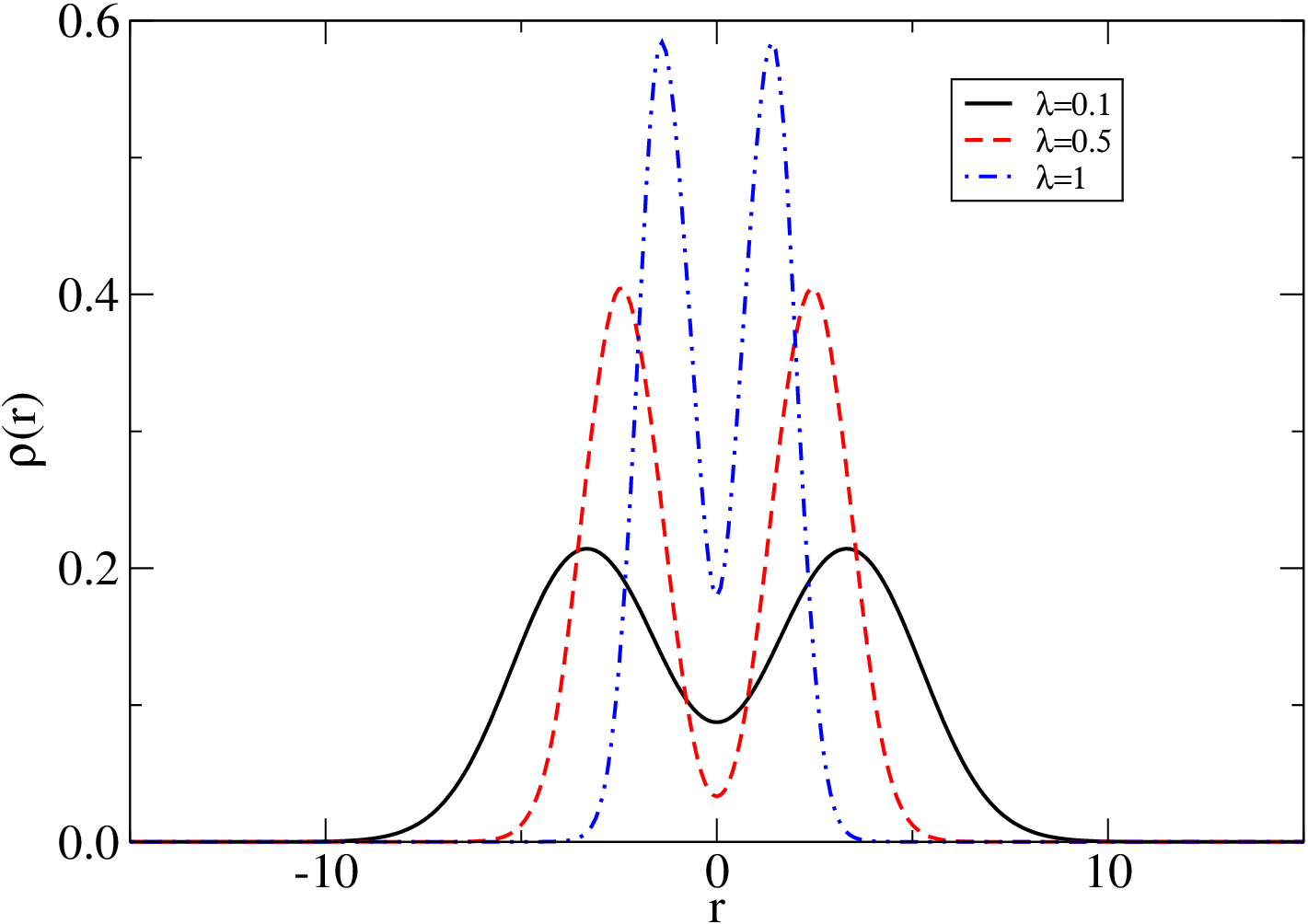}
\end{minipage}\\[1em]
  \caption{Electron density of the two electron $S=1$ system coupled to light. Left $\lambda=0.1$ a.u., right $\omega_p=1$ a.u.}
\label{ph2}
\end{figure}

\begin{figure}
  \begin{minipage}{0.85\textwidth}
        \includegraphics[width=0.35\textwidth]{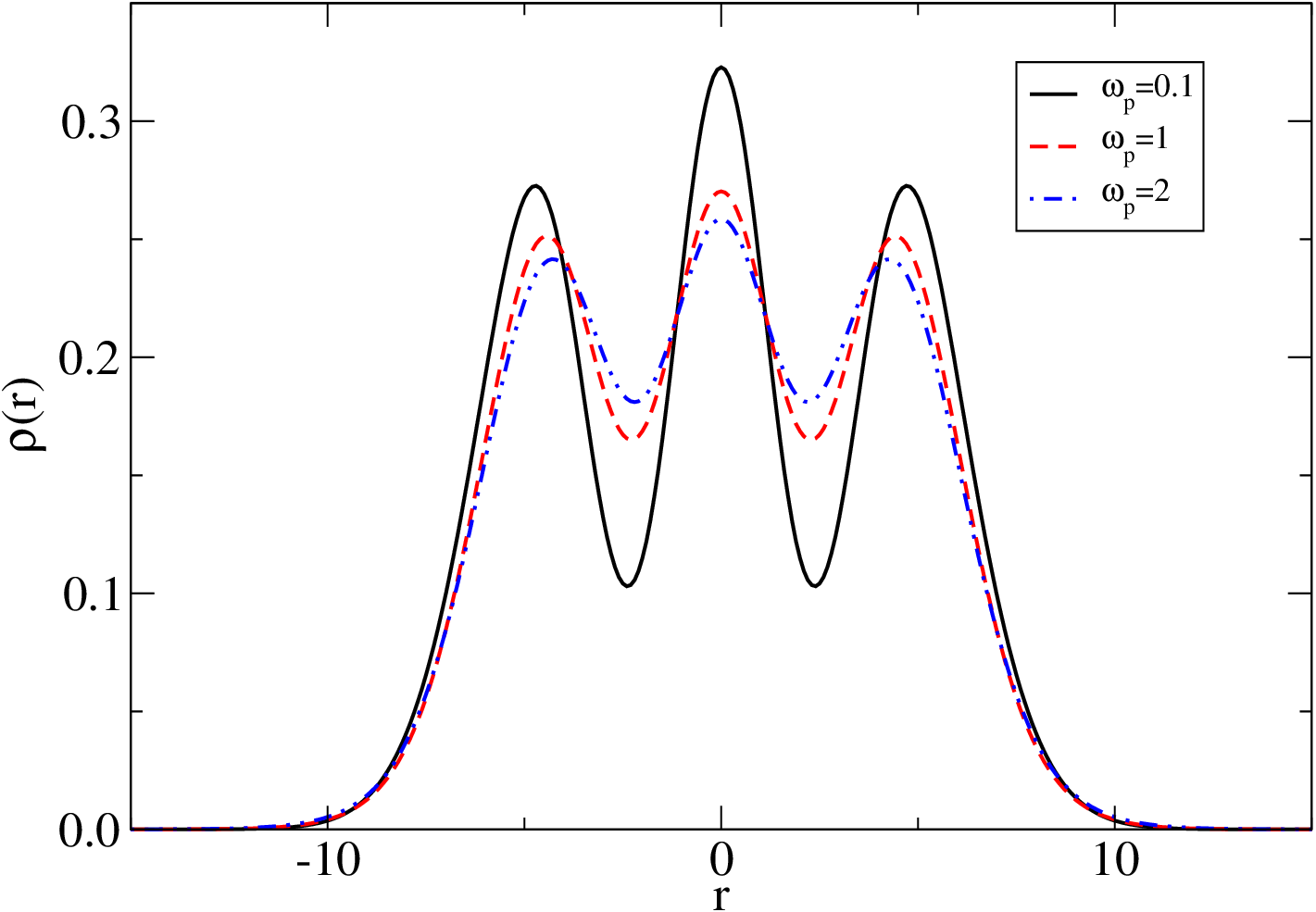}\quad
        \includegraphics[width=0.35\textwidth]{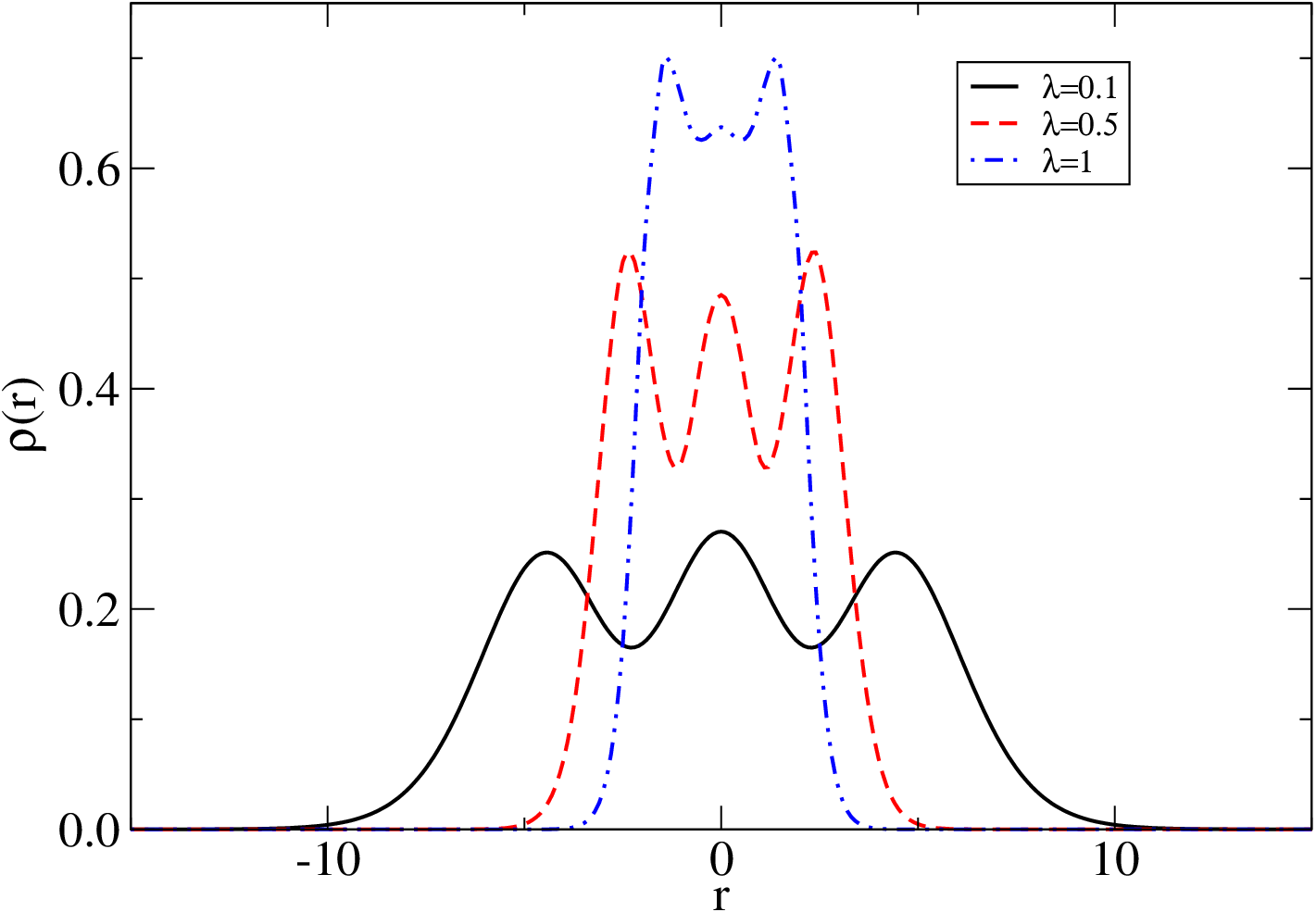}\\\vskip 1cm
        \includegraphics[width=0.35\textwidth]{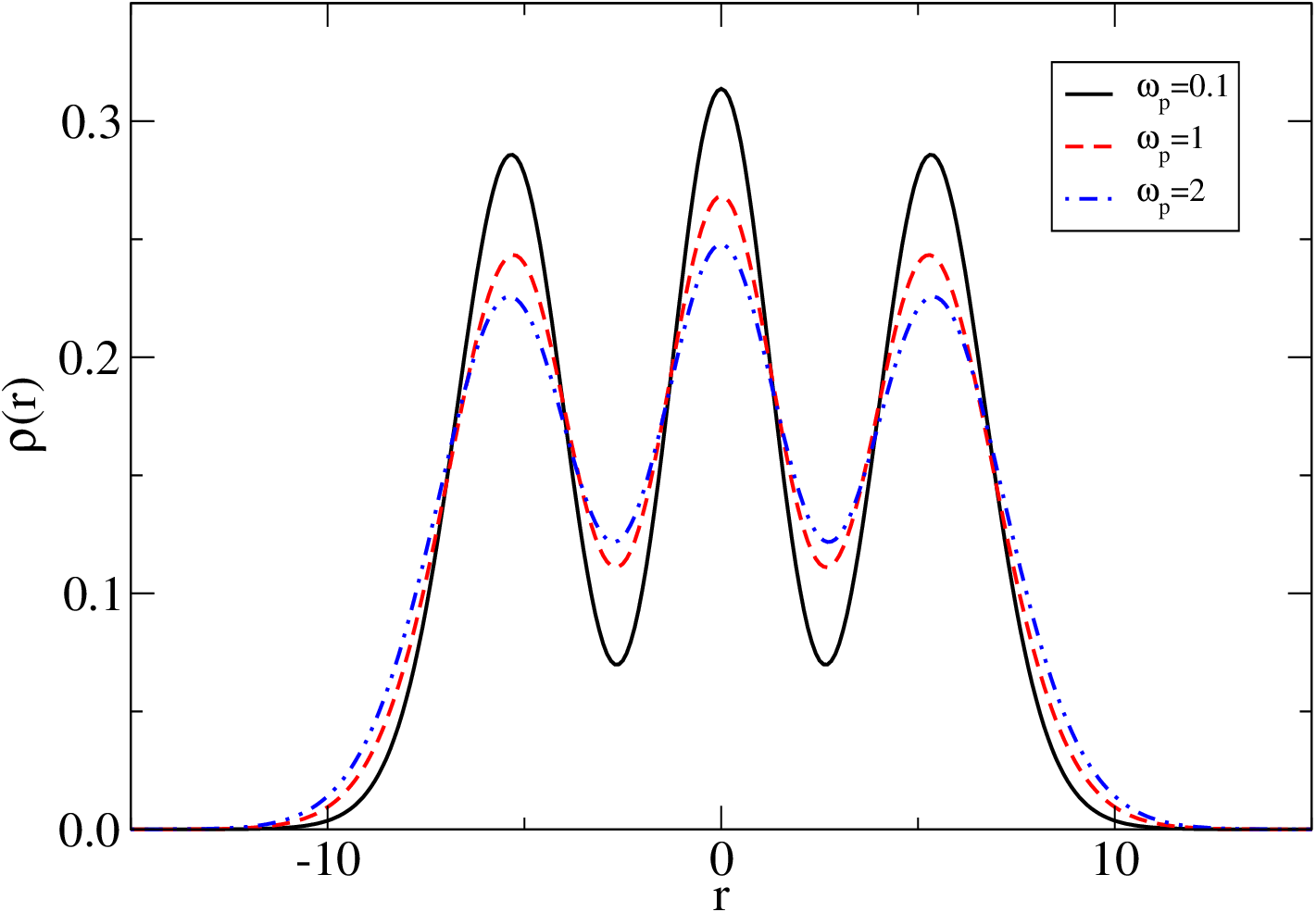}\quad
        \includegraphics[width=0.35\textwidth]{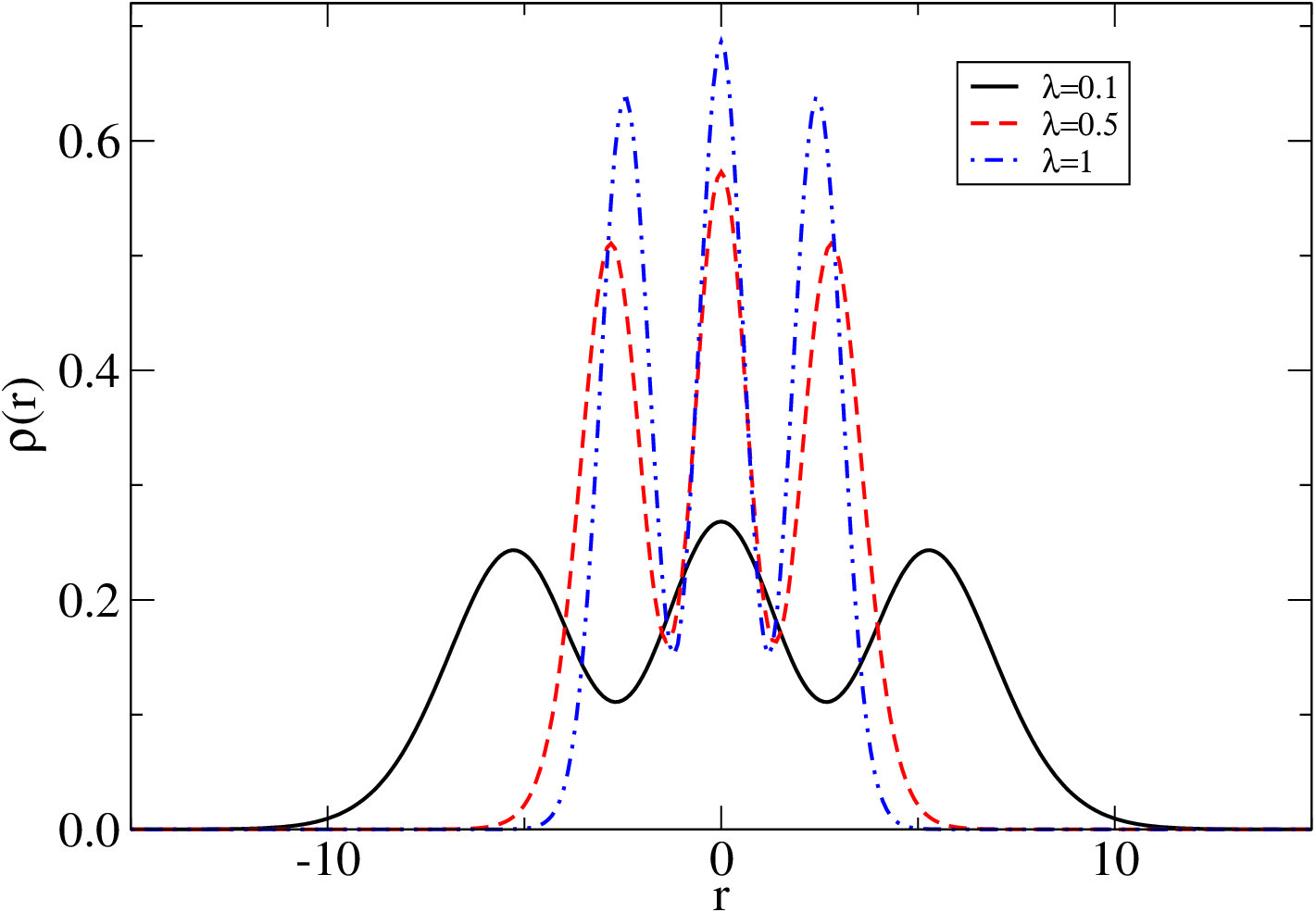}
\end{minipage}\\[1em]
  \caption{Electron density of the three electron system coupled to light. Top $S=1/2$, bottom $S=3/2$; left $\lambda=0.1$ a.u., right $\omega_p=1$ a.u.}
\label{ph3}
\end{figure}

\begin{figure}
  \begin{minipage}{0.85\textwidth}
      \includegraphics[width=0.35\textwidth]{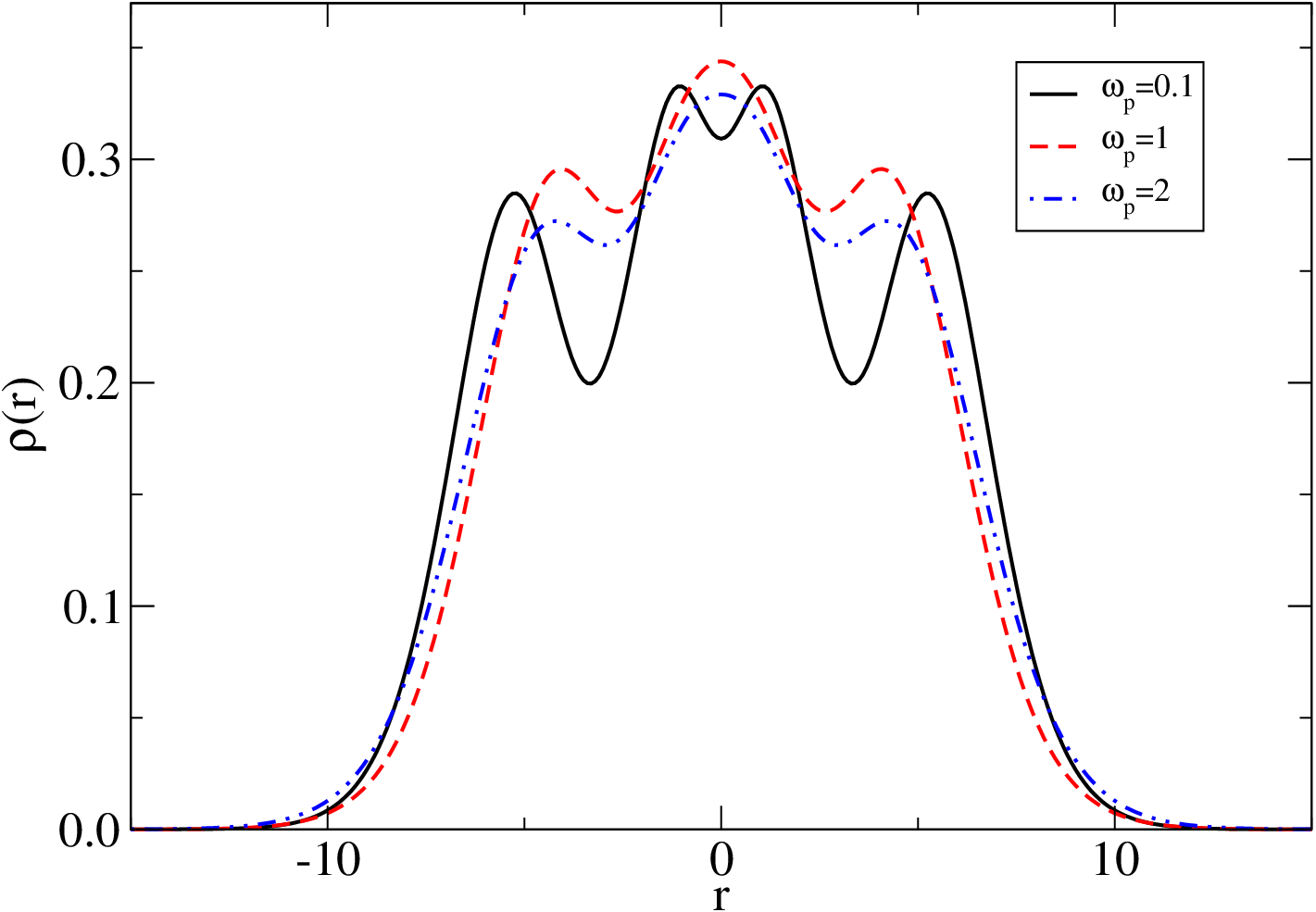}\quad
      \includegraphics[width=0.35\textwidth]{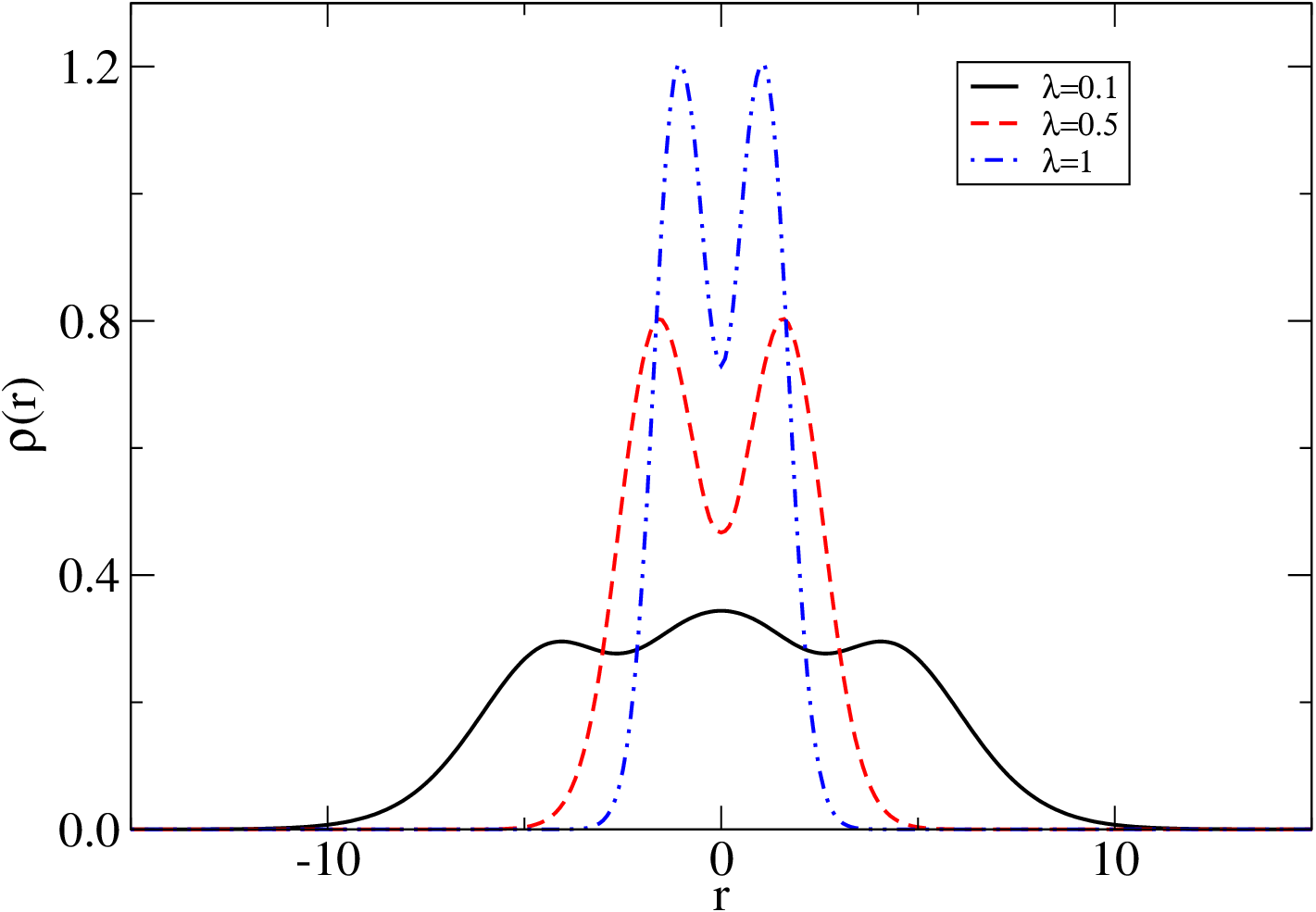}\\\vskip 1cm
        \includegraphics[width=0.35\textwidth]{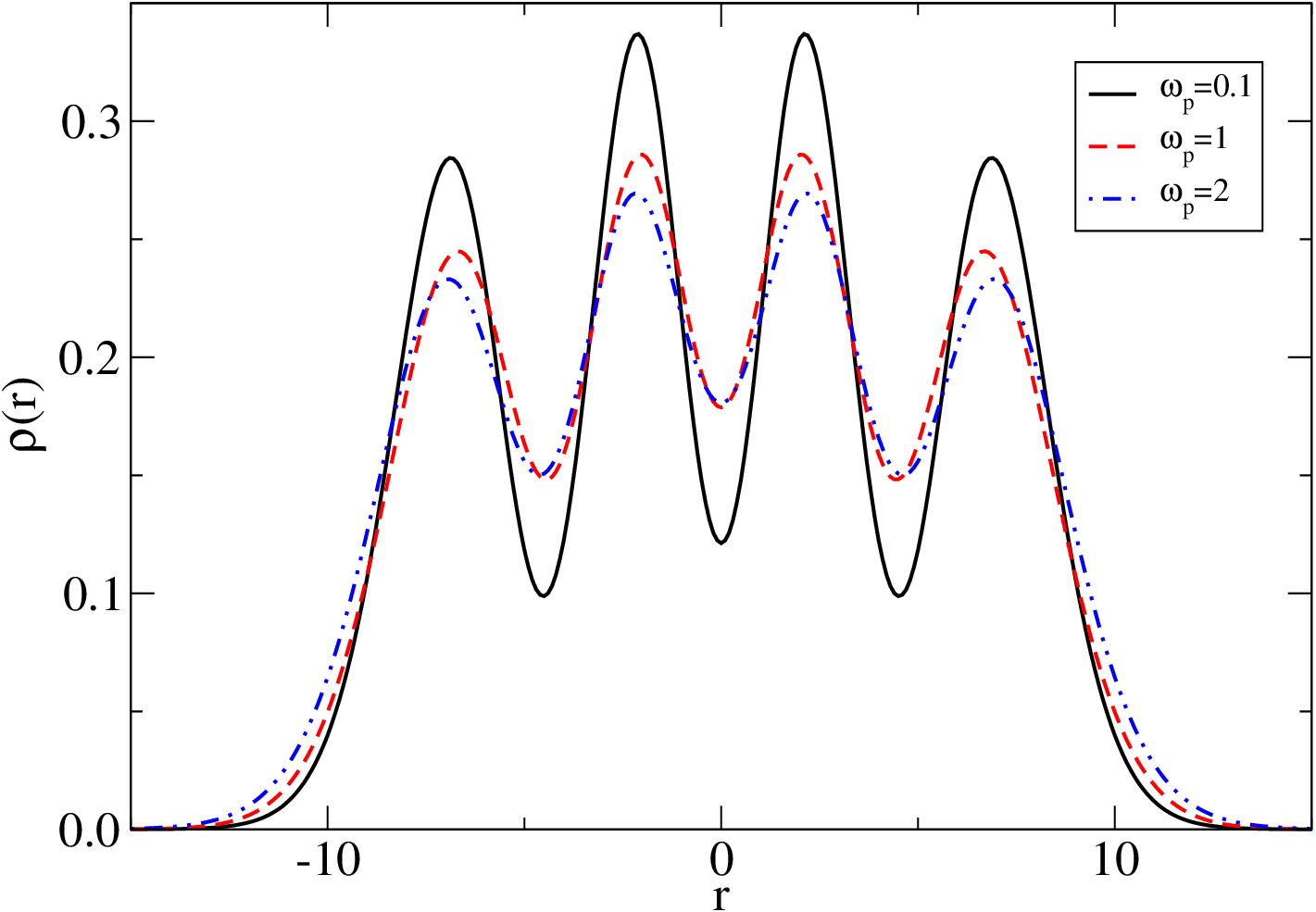}\quad
        \includegraphics[width=0.35\textwidth]{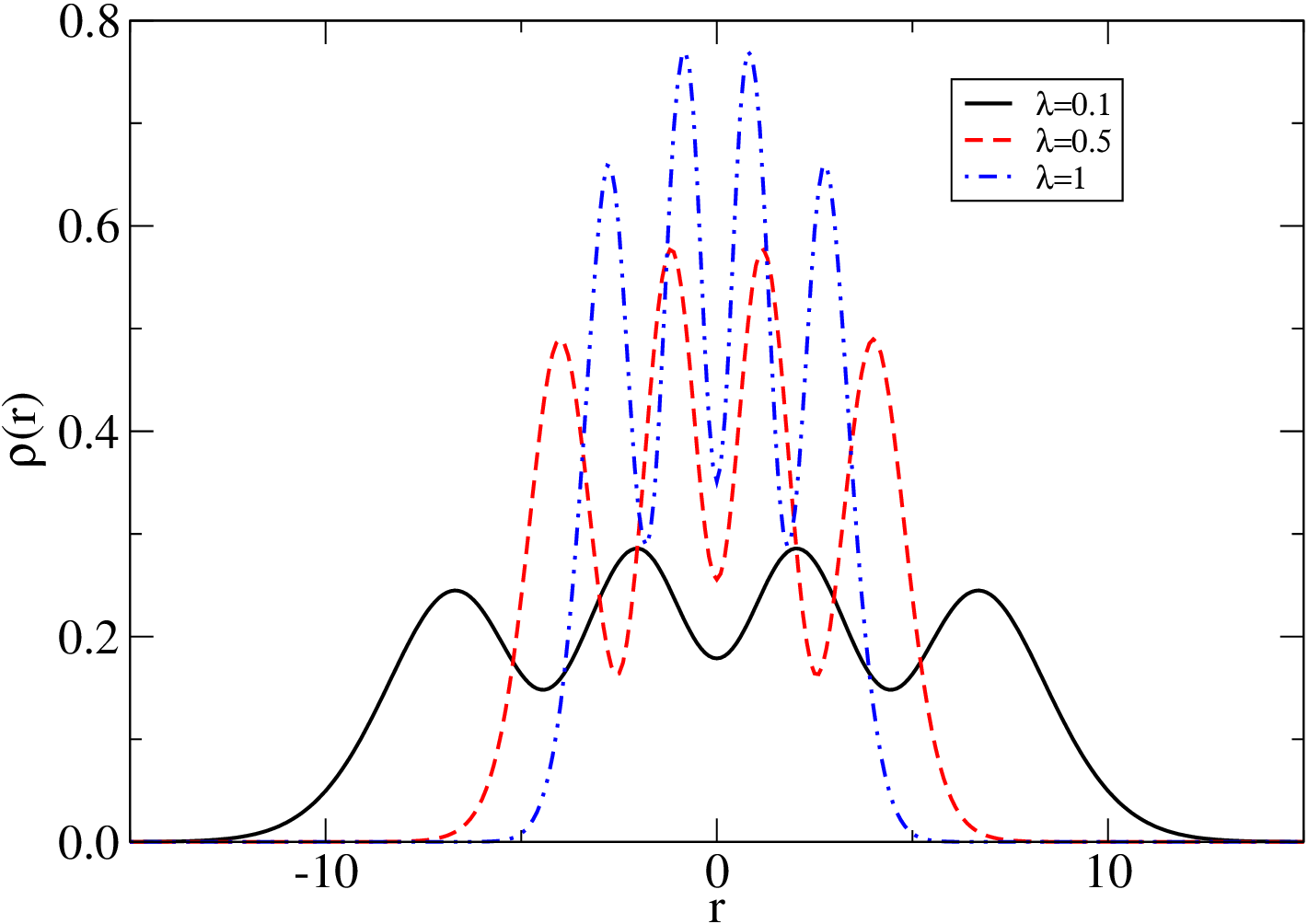}
\end{minipage}\\[1em]
  \caption{Electron density of the four electron system coupled to light. Top $S=0$, bottom $S=2$; left $\lambda=0.1$ a.u., right $\omega_p=1$ a.u.}
\label{ph4}
\end{figure}

In figures \ref{ph2}-\ref{ph4}, we further present our results of ECG 
calculations for N=2-4 electron Wigner crystals formed and controlled  by light-matter coupling. In these systems we use a weak harmonic oscillator confining potential $(\omega=0.1$ a.u.). Although this 
confinement allows the density to spread out far away from the center, the interaction of these systems with light strongly squeezes the density toward the center.

We test the systems for three different $\omega_p$'s (photon frequency) and different coupling strengths. 
The first strong coupling $\lambda=1$ confines the system into a [-5,5] a.u. box; the second 
moderate coupling $\lambda=0.1$ confines the system into a [-10,10] a.u. box; 
and the weakest $\lambda=0.01$ forces the system into a roughly [-12,12] a.u. box. In this case we do not make comparison with DFT
because the LDA based DFT does not produce meaningful results. Only
selected spin states are included, as others show similar density distributions. 

\begin{figure}
  \begin{minipage}{0.85\textwidth}
        \includegraphics[width=0.35\textwidth]{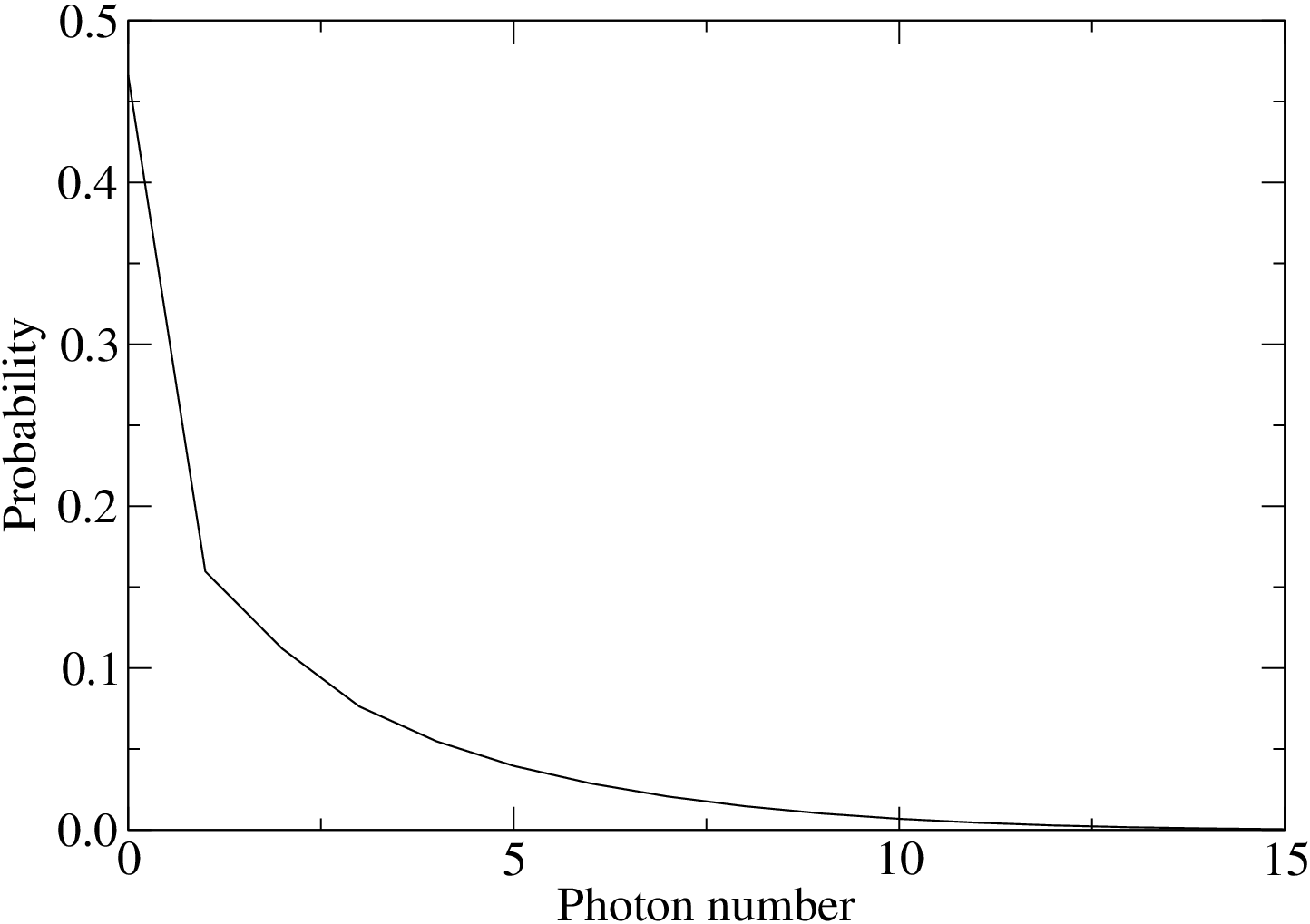}
\end{minipage}\\[1em]
  \caption{Photon number as a function of the photon space for a single electron system with $\lambda=0.1$ a.u., and $\omega_p=0.1$ a.u.}
\label{ph}
\end{figure}

Note that in this case we are not merely dealing with a harmonic 
confinement as in the previous section, but as the wave function in
Eq. \eqref{pwf} shows, the electrons are confined in different photon
number spaces coupled to each other.  The electron density is the sum
of the electron density calculated in the orthogonal photon number
spaces. An example is shown in Fig. \ref{ph} for a case of a single 
electron. The figure shows the probability of different photon number
spaces, the fraction of the norm of the wave function belonging to
different $|n\rangle$ in Eq. \eqref{pwf}. In this one electron case  
the coupling is relatively strong; high photon spaces are coupled and less than fifty percent  of the density is in the zero photon space. 

Fig. \ref{ph2} shows the electron density of an $N=2$ system as a function of $\omega_p$ and $\lambda$. The coupling between  different
photon spaces is controlled by $g$ (see Eq. \eqref{str}), and the
strength of confinement in a given photon space is determined by
$\lambda$.  For a given $\lambda$ value the dependence on $\omega_p$ 
is relatively small. For a given $\omega_p$ the positions and structures
of the peaks are strongly dependent on $\lambda$. One significant
difference between the harmonic confinement and the photon coupled case
(Figs. \ref{2ed} and \ref{ph2}) is that the density is much smaller
between the peaks in the latter case. 

Figs. \ref{ph3} and \ref{ph4} show a similar dependence 
on $\omega_p$ for a given $\lambda$. The confinement is determined by
$\lambda$ and the density distributions have almost identical widths
and peak positions. By increasing $\omega_p$, the peak structure may
become less emphasized for non spin-polarized cases (E.g. for $N=4$ $S=0$ case only two or three peaks manifest for larger $\omega_p$ or $\lambda$).  Similar to harmonic confinement, 
the number of the density peaks still 
matches with the number of electrons in spin-polarized cases.

For a given $\omega_p$, the dependence on $\lambda$ is strong
(Figs. \ref{ph2}-\ref{ph4}). Larger $\lambda$ values make more
compact systems. Overall, the $\lambda$ dependence seems to be very 
similar in all cases. Photon spaces with small photon number
($n$=0,1,2) contain almost all the electron densities even for stronger $\lambda$. The dependence of the densities and energies \cite{Supp}
on the photon frequency is moderate. The strong dependence on $\lambda$
is due to two reasons. First, as Eq. \ref{str} shows the coupling is proportional to $\lambda$. Second and more importantly the dipole
self-interaction strength grows as $\lambda^2$. The latter fact also explains that for a given $\lambda$ and changing $\omega$ (see Figs.
\ref{ph2},\ref{ph3} and \ref{ph4}), the spread of the density is nearly
identical and only the relative heights of the peaks change. The dipole
self-interaction term is responsible for squeezing the density toward
the center. 

\begin{figure}
 \includegraphics[width=0.5\textwidth]{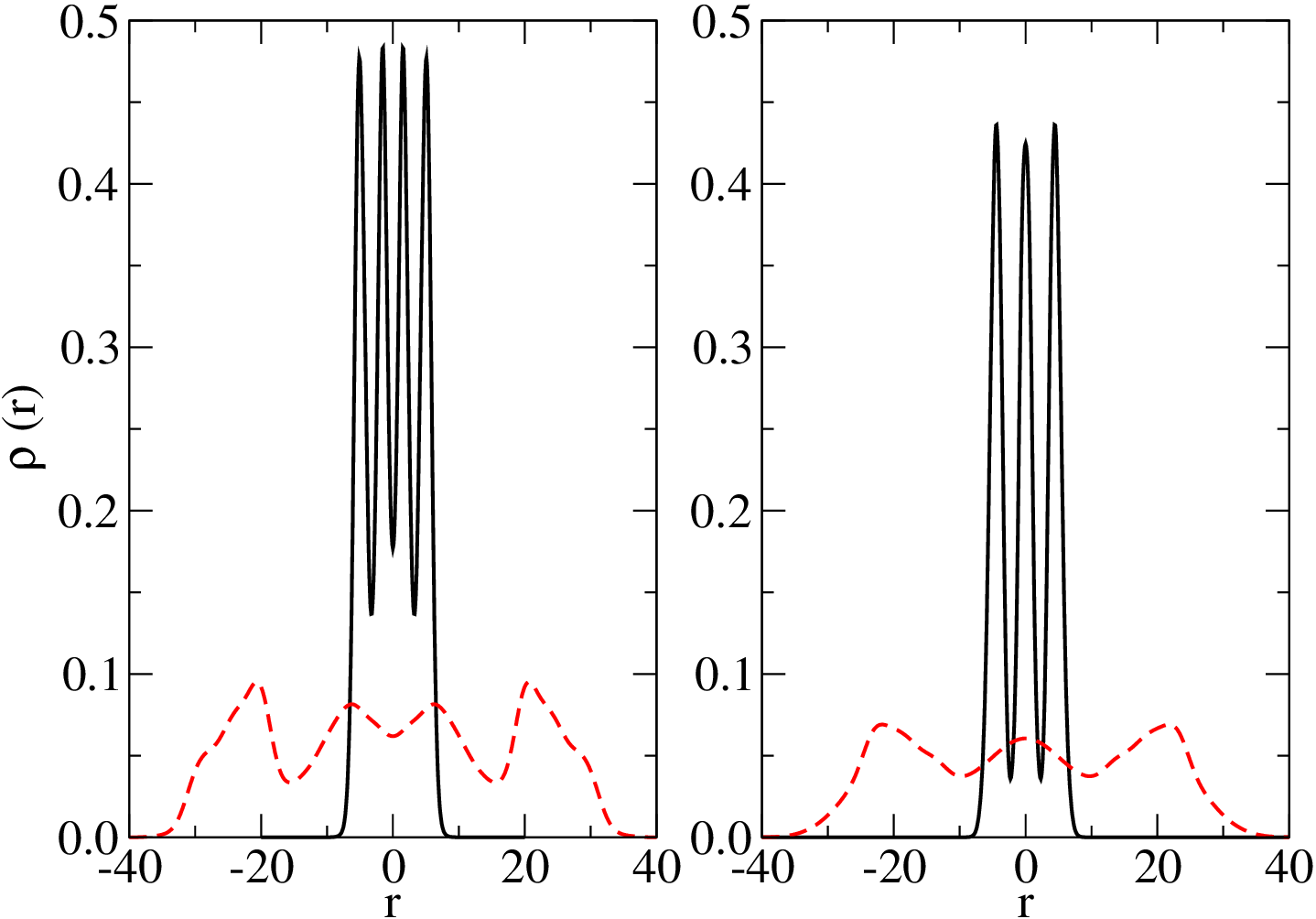}
 \caption{Electron density of the four (left) and three (right) electron 
 systems coupled to light. The solid line shows the density 
 for $\lambda=1.$ a.u., $\omega_p=0.5$ a.u. and $\omega=0.001$ a.u., 
 the dashed line shows the density without coupling.}
\label{ph5}
\end{figure}

These systems would not be bounded harmonically without an external  confining potential. The photons couple to the electrons through
the center of mass coordinate of the system (see Eqs. \eqref{epi} and 
\eqref{dpi}). The total wave function of the electrons can be factorized as a wave function of relative motion (depending on the
relative coordinate) and the wave function of the center of mass motion (depending on $X$ only). If there is no confinement, then the relative motion is governed by the repulsive Coulomb interaction 
and the system dissociates. The strength of the confinement, however, plays very little role. Fig. \ref{ph5} shows three and four electron 
systems with a very weak confining potential for the spin polarized 
(S=3/2 and S=2) cases. Without coupling, the density spreads out to 40 a.u. The coupling squeezes the density and the electrons form a  tightly localized  Wigner crystal. These three and four electron 
densities are very similar to the $\omega=0.1$ a.u. cases shown in
Figs. \ref{ph3} and \ref{ph4}. 

\section{Summary}
1D few-electron systems are investigated using ECG basis functions.
All matrix elements are analytically calculated and the basis parameters
are optimized to generate flexible basis and accurate wave functions.
N=2-6 electron systems with different spin states are studied.

Two different confinements are considered. In the first case, an external potential is used to localize the electrons. In the second case, there is a weak confining potential but the electrons are coupled to light and the dipole self-polarization determines the
confinement. 1D Wigner crystal-like structures appear in both cases and
there is a similar tendency in the shape of the density as the confinement strength changes. 

In spin-polarized cases the number of density peaks is equal to the
number of electrons because the shell structure, created by the
confining potential, dominates. For non spin-polarized cases the
number of peaks depends on the confinement strength and  the total
spin. 

We have shown that the Wigner crystal structure is not suppressed by strong confinement. In the strong confinement regime, the Coulomb interaction 
becomes negligible compared to the kinetic energy and the confinement.
But in this region in 1D a shell structure is overwhelming and the
nodes of the wave functions define the crystal-like peaks in the density.

Simpler models like DFT based calculations can capture the Wigner
crystal structure in certain cases, especially for spin-polarized
systems. Better exchange-correlation potentials can potentially extend
the range of applications of the DFT based approach to other cases. 
The advantage of the DFT is that it is easily applicable to much
larger electron systems. The densities calculated by ECG can be
used to create better exchange-correlation potentials for these 1D
systems.

We have considered electrons with (soft) Coulomb interaction in this work, but other systems with repulsive interactions, such as degenerate Fermi gases in confinements or cavities\cite{PhysRevA.83.043606}, would be expected to show similar structures.

Wigner crystals in systems confined by external potentials have already been observed 
\cite{Shapir870}. The experimental realization of the light coupled systems might be possible by using nanotubes or optical lattices in cavities.

\begin{table}
\begin{tabular}{|c|c|c|c|}
 \hline

    & $\omega$ & $E$\,(ECG) & $E$\,(DFT)\\
 \hline
 \multicolumn{4}{|c|}{$2e^-$}\\
 \hline
S=0& 0.1       & 0.392 &  0.005\\
   & 1.0       & 1.774  &  1.111\\
 \hline
S=1 & 0.1       & 0.396  & -0.1 \\
   & 1.0       & 2.554  & 1.827  \\
\hline
\multicolumn{4}{|c|}{$3e^-$}\\
\hline
S=0.5 & 0.1       & 1.009& 0.256\\
   & 1.0       & 4.481& 3.385\\
 \hline
S=1.5& 0.1       & 1.016  & 0.246\\
   & 1.0       & 6.078   & 4.872  \\ 
 \hline
 \multicolumn{4}{|c|}{$4e^-$}\\
 \hline
S=0 & 0.1       & 1.877  & 0.982\\
   & 1.0       & 7.808  & 6.261\\
 \hline
S=1& 0.1       & 1.887  & 0.846\\
   & 1.0       & 8.589  &  7.005 \\ 
 \hline
 S=2& 0.1       & 1.894  & 0.837\\
   & 1.0       & 11.024 & 9.293  \\ 
 \hline
 \multicolumn{4}{|c|}{$5e^-$}\\
 \hline
S=0.5 & 0.1       & 2.999 & 1.678\\
   & 1.0       & 12.490 & 10.443\\
 \hline
S=1.5& 0.1       & 2.985 &  1.671\\
   & 1.0       & 14.069  &   11.955\\ 
 \hline
 S=2.5& 0.1       & 3.020 & 1.663\\
   & 1.0       & 17.379 &  15.064\\ 
   \hline
   \multicolumn{4}{|c|}{$6e^-$}\\
   \hline
S=0 & 0.1       & 4.362 &  2.822\\
   & 1.0       & 17.733 &  15.164\\
 \hline
S=1 & 0.1       & 4.357 & 2.715\\
   & 1.0       & 18.566 & 15.919 \\ 
 \hline
 S=2& 0.1       & 4.336 &  2.716\\
   & 1.0       & 20.911 &   18.221\\ 
 \hline
 S=3 & 0.1       & 4.413 &  2.716\\
   & 1.0       & 25.099 &  22.167 \\ 
 \hline
\end{tabular}
\caption{Total energy $E$ (in atomic units) for few-electron systems as a function of the external confinement strength $\omega$.}
\label{t2}
\end{table}

\onecolumngrid
\clearpage

\begin{acknowledgments}
This work has been supported by the National Science
Foundation (NSF) under Grant No. IRES 1826917.
\end{acknowledgments}


\end{document}